\documentclass[11pt]{article}
\usepackage{amsmath}
\usepackage{amsfonts,amssymb}
\usepackage{mathrsfs}
\usepackage{diagrams}
\usepackage{oldgerm}
\usepackage{psfrag}
\usepackage{epsfig}

\newcommand{\N}{\mathbb{N}}	      
\newcommand{\Rl}{\mathbb{R}}
\newcommand{\Cl}{\mathbb{C}}
 
\newcommand{\Hil}{\mathcal{H}}
\newcommand{\VV}{\mathcal{V}}

\newcommand{\DD}{\mathcal{D}} 
 
\newcommand{\Ss}{\mathscr{S}}   		
\newcommand{\OO}{O}   				
\newcommand{\PGpo}{\mathcal{P}_+^\uparrow}   	
\newcommand{\PG}{\mathcal{P}} 
\newcommand{\OOO}{\mathscr{O}} 
\newcommand{\Di}{\mathscr{D}} 
\newcommand{\NN}{\mathcal{N}}  
\newcommand{\X}{\mathcal{X}}  
\newcommand{\Y}{\mathcal{Y}} 
\newcommand{\CC}{\mathscr{C}}			
\newcommand{\SF}{\mathcal{S}}
\newcommand{\frS}{\textfrak{S}} 		
\newcommand{\Tu}{\mathcal{T}}
\newcommand{\Ba}{\mathcal{B}}
\newcommand{\Cu}{\mathcal{C}}  			

\newcommand{\A}{\mathcal{A}}
\newcommand{\B}{\mathcal{B}}
\newcommand{\M}{\mathcal{M}}

\def\bte{{\mbox{\boldmath{$\te$}}}}
\def\bla{{\mbox{\boldmath{$\lambda$}}}}
\def\bze{{\mbox{\boldmath{$\zeta$}}}}
\def\bxi{{\mbox{\boldmath{$\xi$}}}}
\def\beps{{\mbox{\boldmath{$\varepsilon$}}}}

\def\bof{{\mbox{\boldmath{$f$}}}}
\def\bk{{\mbox{\boldmath{$k$}}}}
\def\bl{{\mbox{\boldmath{$l$}}}}

\newcommand{\bno}[1]{|\!|\!|#1|\!|\!|}		

\def\br{{\mbox{\boldmath{$r$}}}}

\newcommand{\sbte}{{\mbox{\scriptsize \boldmath $\theta$}}}

\newcommand{\sbk}{{\mbox{\scriptsize \boldmath $k$}}}
\newcommand{\sbl}{{\mbox{\scriptsize \boldmath $l$}}}

\newcommand{\sbla}{{\mbox{\scriptsize \boldmath $\lambda$}}}
\newcommand{\sbeps}{{\mbox{\scriptsize \boldmath $\varepsilon$}}}
\newcommand{\sbr}{{\mbox{\scriptsize \boldmath $r$}}}

\newcommand{\supp}{\mathrm{supp}\,}

\newcommand{\te}{\theta}
\newcommand{\la}{\lambda}
\newcommand{\La}{\Lambda}
\newcommand{\eps}{\varepsilon}
\newcommand{\Om}{\Omega}

\newcommand{\us}{\underline{s}}
\newcommand{\fti}{\widetilde{f}}
\newcommand{\ghat}{\hat{g}}
\newcommand{\fhat}{\hat{f}}

\newcommand{\zd}{z^{\dagger}}

\newcommand{\iin}{_\mathrm{in}}
\newcommand{\ex}{_\mathrm{ex}}
\newcommand{\oout}{_\mathrm{out}}

\newtheorem{theorem}{Theorem}[section]
\newtheorem{proposition}[theorem]{Proposition}
\newtheorem{definition}[theorem]{Definition}
\newtheorem{lemma}[theorem]{Lemma}
\newtheorem{app-lemma}{Lemma}[subsection]

\newenvironment{proof}{\medskip \noindent {\em Proof:}}{\hfill
  $\square$ \\[2mm] \indent}
\newenvironment{acknowledgements}{\bigskip\bigskip\noindent\small {\bf Acknowledgements:}}

\newlength{\dinwidth}
\newlength{\dinmargin}
\begin{document}

\title{Construction of Quantum Field Theories\\ with Factorizing S-Matrices}

\author{Gandalf Lechner\\[4mm]
 	\small{Erwin Schr\"odinger Institute for Mathematical Physics},\\
 	\small{Boltzmanngasse 9, A-1090 Vienna, Austria}\\[4mm]
 	\small email: {\tt gandalf.lechner@esi.ac.at}
}

\date{February 12, 2007}
\maketitle

\begin{abstract}
A new approach to the construction of interacting quantum field theories on two-dimensional Minkowski space is discussed. In this program, models are obtained from a prescribed factorizing S-matrix in two steps. At first, quantum fields which are localized in infinitely extended, wedge-shaped regions of Minkowski space are constructed explicitly. In the second step, local observables are analyzed with operator-algebraic techniques, in particular by using the modular nuclearity condition of Buchholz, d'Antoni and Longo.

Besides a model-independent result regarding the Reeh-Schlieder property of the vacuum in this framework, an infinite class of quantum field theoretic models with non-trivial interaction is constructed. This construction completes a program initiated by Schroer in a large family of theories, a particular example being the Sinh-Gordon model. The crucial problem of establishing the existence of local observables in these models is solved by verifying the modular nuclearity condition, which here amounts to a condition on analytic properties of form factors of observables localized in wedge regions.

It is shown that the constructed models solve the inverse scattering problem for the considered class of S-matrices. Moreover, a proof of asymptotic completeness is obtained by explicitly computing total sets of scattering states. The structure of these collision states is found to be in agreement with the heuristic formulae underlying the Zamolodchikov-Faddeev algebra.
\end{abstract}

\section{Introduction}
In relativistic quantum field theory, the rigorous construction of models with non-trivial interaction is still a largely open problem. Apart from the well-known results of Glimm and Jaffe \cite{glimmjaffe}, most interacting quantum field theories are treated only perturbatively, usually without any control over the perturbation series.

The main difficulties in the construction of interacting quantum field theories arise from the principle of Einstein causality, demanding that physical observables must be strictly {\em local}, i.e. represented by commuting operators when spacelike separated. In view of this problem, several authors \cite{BuLe,BuSuAdS,S1,SW,BCFT,MSY} have proposed to construct models by first considering easier manageable, non-local theories, and then passing to a local formulation in a second step.

A particular example of such a constructive scheme is the program initiated by Schroer  \cite{SW,S1,S2}, which deals with quantum field theories on two-dimensional Min\-kowski space. Here the interaction of the models to be constructed is not formulated in terms of classical Lagrangians, but rather by prescribed S-matrices, i.e. the inverse scattering problem is considered. The S-matrices are assumed to be {\em factorizing} \cite{iagol}, i.e. of the type found in completely integrable models such as the Sinh-Gordon-, $O(N)$ Sigma-, or Thirring model \cite{abdalla,dorey}.

Adopting the idea of constructing local theories by first considering non-local auxiliary quantities, one starts in this program from a given factorizing S-matrix $S$ and defines two non-local quantum fields $\phi$, $\phi'$ depending on $S$ \cite{gl1}. Although these fields are not local, they are {\em relatively wedge-local} to each other in the following sense. Consider the so-called {\em right wedge}
\begin{equation}\label{def:WR}
 W_R:=\{x\in\Rl^2\,:\,x_1>|x_0|\}\,,
\end{equation}
and its causal complement $W_L:=W_R'=-W_R$, the {\em left wedge}. Then $\phi(x)$ and $\phi'(y)$ commute (in a suitable sense) if the wedges $W_L+x$ and $W_R+y$ are spacelike separated. Hence $\phi(x)$ is {\em not}, as usual, localized at the spacetime point $x$, but rather spread out over the infinitely extended spacetime region $W_L+x$.

The advantage of these non-local field operators is that they can be cast into a very simple form in momentum space. In fact, the only difference to free fields are the deformed commutation relations of their creation and annihilation parts, which form a representation of the Zamolodchikov-Faddeev algebra \cite{ZZ}. In particular, $\phi$ and $\phi'$ create only single particle states from the vacuum, without accompanying vacuum polarization clouds. In view of this special property, such operators have been termed {\em polarization-free generators} \cite{S2}, see \cite{BBS} for a model-independent discussion of this concept.

The first step of the inverse scattering construction of models with factorizing S-matrices, i.e. the construction and analysis of their wedge-local fields, has by now been completed for a large class of underlying scattering operators \cite{gl1,SW,BuLe}. It is the aim of the present article to accomplish the second step of the construction, i.e. the passage from the wedge-local fields to theories complying with the principle of locality, for the family of S-matrices considered in \cite{gl1}.

Usually the task of classifying and constructing quantum field theories with factorizing S-matrices is taken up in the so-called form factor program \cite{smirnov,kb2}. In that approach, one studies local fields $A$ in terms of their matrix elements in scattering states (form factors). The $n$-point functions of $A$ are then represented as infinite series of integrals over form factors. These form factors have been calculated for a multitude of models \cite{kb1,ff:fms,ff:sigma,ff:Zn}, at least for the lowest particle numbers. But in almost all cases\footnote{The only non-trivial example of a proof of convergence of a form factor expansion which is known to us is the case of a two-point function in the Yang-Lee model (F.~A.~Smirnov, private communication).} one is still lacking control over the series representing the $n$-point functions  \cite{kb2}. This is due to the complicated form factor functions, and can be understood as a consequence of the complicated momentum space structure a local quantum field must have in the presence of interaction. So at present, the existence of models with prescribed S-matrices cannot be decided within the form factor program.
 
If a given collision operator is expected to be related to a classical Lagrangian, there is also the possibility of applying the Euclidian techniques of constructive quantum field theory. Along these lines, the existence of the Sine-Gordon model and the non-triviality of its S-matrix have been established by Fr\"ohlich \cite{fr1,fr2}. But the hard task of explicitly computing the scattering operator and making contact with the form factor approach is still an open problem in this framework.

In the context of the wedge-local fields $\phi$, $\phi'$, local observables can be characterized by commutation relations with $\phi$ and $\phi'$ \cite{SW}. Solving these relations amounts to solving the form factor program. We will follow here a different approach, motivated by the observation that for the analysis of basic questions, such as the existence of theories with certain properties, it is not necessary to have explicit expressions for strictly local quantities. The problem to decide if a given factorizing S-matrix is realized as the collision operator of a well-defined quantum field theory can be solved by considering only the structure of observables localized in wedge regions.

To accomplish this task, one has to answer the question whether the wedge-local models defined by the fields $\phi$, $\phi'$ contain also observables localized in bounded spacetime regions. A strategy how to solve this existence problem was proposed in \cite{BuLe}. The main idea is to consider the algebras generated by bounded functions of $\phi$, $\phi'$ rather than the fields themselves. One proceeds to a net of {\em wedge algebras}, i.e. a collection of von Neumann algebras $\A(W)$, where $W$ runs through the family $\{W_R+x,W_L+x\,:\,x\in\Rl^2\}$ of all wedges in two-dimensional Minkowski space. In this formulation, powerful operator-algebraic techniques become available for the solution of the existence problem, which have not been employed in other approaches. 

It has been shown in \cite{BuLe} that non-trivial observables localized in a double cone region of the form $W_R\cap(W_L+x)$, $x\in W_R$ (cf. figure \ref{fig:dc}, p. \pageref{fig:dc}) do exist if the so-called {\em modular nuclearity condition} \cite{BDL1} holds, i.e. if the map
\begin{equation}
 \Xi(x):\A(W_R)\longrightarrow\Hil,\qquad \Xi(x)A:=\Delta^{1/4}U(x)A\Om,
\end{equation}
is nuclear. Here $\Delta$ denotes the modular operator \cite{KadRin} of $(\A(W_R),\Om)$. So the existence of {\em local} observables can be established by estimates on {\em wedge-local} quantities. 

In the context of theories with factorizing S-matrices, the crucial question arises whether the modular nuclearity condition holds in such models. We will show here that this condition takes a very concrete form in these models, and can be solved by analyzing analytic continuations of form factors of observables localized in wedges. As our main result, we will give a proof of the nuclearity condition for a large class of underlying S-matrices, thereby establishing the existence of the corresponding models as well-defined, local quantum field theories. Moreover, once the modular nuclearity condition has been established, it is possible to apply the usual methods of scattering theory. Doing so, we will compute total sets of $n$-particle collision states in these models, and prove that the construction solves the inverse scattering problem.
\\
\\
This article is organized as follows. In Section 2, we extend the analysis of \cite{BuLe} and derive further consequences of the modular nuclearity condition in a model-independent, operator-algebraic framework. It is shown there that the Reeh-Schlieder property of the vacuum \cite{streater}, which is a prerequisite for doing scattering theory \cite{araki}, follows from the nuclearity condition (Theorem \ref{thm2}).

For the sake of self-containedness, the basic definitions and results regarding the models based on the fields $\phi,\phi'$ are recalled in Section \ref{sec:model}. Also the classes of factorizing S-matrices which we consider are defined there (Definitions \ref{def:S2} and \ref{def:S0}).

In Section 4, analytic continuations of form factors of observables localized in wedges are studied. It is then shown in Section 5 how these analytic properties can be used to verify the modular nuclearity condition. We obtain two proofs of different generality (Theorems \ref{thm:distalsplit} and \ref{thm:-1nuclearity}).

Section 6 is devoted to a study of the collision states of the constructed models. It is shown that our construction solves the inverse scattering problem for the considered class of S-matrices (Theorem \ref{thm:S-matrix}), and a proof of asymptotic completeness is given (Proposition \ref{prop:ac}).

The paper closes in Section 7 with our conclusions, and the technical proof of a lemma needed in Section 4 can be found in the appendix.
\\
\\
This article is based on the PhD thesis of the present author \cite{gl-phd}.

\section{Construction of Local Nets from a Wedge Algebra}\label{sec:alg}
In this section we discuss some model-independent aspects of the construction procedure. In contrast to the following chapters, we will here base our analysis only on assumptions which are satisfied in a wide class of quantum field theories, and do not use the special structure of the integrable models to be studied later.

As explained in the Introduction, it is our aim to construct strictly local quantum field theories, but use auxiliary objects which are localized only in wedge regions during the construction. We therefore consider an algebra $\M$ modelling the observables localized in the reference wedge $W_R$ \eqref{def:WR}, and a representation $U$ of the two-dimensional translation group $(\Rl^2,+)$. Given these data, we will construct a corresponding quantum field theory by specifying, for arbitrary regions $\OO$ in two-dimensional Minkowski space, the algebras $\A(\OO)$ containing all its observables localized in $\OO$, and show that they have the right physical properties.

The ``wedge algebra'' $\M$ is taken to be a von Neumann algebra acting on a Hilbert space $\Hil$, which in order to exclude trivialities we assume to satisfy ${\rm dim}\Hil>1$. Moreover, we require
\begin{itemize}
\item[A1)] $U$ is strongly continuous and unitary. The joint spectrum of the generators $P_0$, $P_1$ of $U(\Rl^2)$ is contained in the forward light cone $\{p\in\Rl^2:p_0\geq|p_1|\}$. There is an up to a phase unique unit vector  $\Om\in\Hil$ which is invariant under the action of $U$.
\item[A2)] $\Om$ is cyclic and separating for $\M$.
\item[A3)] For each $x\in W_R$, the adjoint action of the translation $U(x)$ induces endomorphisms on $\M$,
\begin{align}\label{def:Mx}
	\M(x) := U(x)\M U(x)^{-1} \subset \M,\qquad x\in W_R\,.
\end{align}
\end{itemize}
Assumption A1) is standard in quantum field theory \cite{streater,haag}, and identifies $\Om$ as the vacuum vector. A2) and A3) are abstract characterizations of $\M$ as an algebra of observables localized in $W_R$ \cite{borchers-2d,BuLe}.

It should be mentioned that the assumptions A1)-A3) put strict constraints on the algebraic structure of $\M$. The following result has been found in \cite{Longo,driessler}.
\begin{lemma}\label{lem:m31}
 Consider a triple $(\M,U,\Hil)$ satisfying the assumptions A1)-A3). Then $\M$ is a type III$_1$ factor according to the classification of Connes \cite{connes}.
\end{lemma}

Keeping this structure of $\M$ in mind, let us recall how a net $\OO\mapsto\A(\OO)$ of local algebras can be constructed from the data $(\M,U,\Hil)$  \cite{borchers-2d,BuLe}.

As the operators in $\M(x)$ \eqref{def:Mx} are interpreted as observables localized in the translated right wedge $W_R+x$, the elements of the commutant $\M(x)'$ correspond to observables in its causal complement $W_L+x$, where $W_L:=W_R'=-W_R$ is the left wedge. For an operator $A$ representing an observable in the double cone region $\OO_{x,y}$ with vertices $x,y$,
\begin{align}\label{def:dc}
 \OO_{x,y}	&:=	(W_R+x)\cap(W_L+y)\,,\qquad y-x\in W_R,
\end{align}
\begin{figure}[here]
  \begin{center}
    \leavevmode
    \setlength{\unitlength}{1cm}
    \begin{picture}(5,3.2)%
        \centering\epsfig{file=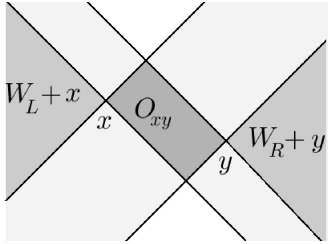,width=4.3cm}
    \end{picture}
  \end{center}
  \caption{The double cone $\OO_{x,y}$ \eqref{def:dc} and its causal complement ${\OO_{x,y}}'=(W_L+x)\cup(W_R+y)$.}
  \label{fig:dc}
\end{figure}
Einstein causality demands that $A$ must commute with both algebras, $\M(x)'$ and $\M(y)$ (see figure \ref{fig:dc}). The maximal von Neumann algebra of operators $A\in\B(\Hil)$ compatible with this condition is 
\begin{align}
 \A(\OO_{x,y})	&:= \M(x) \cap \M(y)'\,.\label{def:AO}
\end{align}
Denoting the set of all double cones in $\Rl^2$ by $\OOO:=\{\OO_{x,y}\,:\,y-x\in W_R\}$, this definition is extended to arbitrary regions $R\subset\Rl^2$ by additivity, 
\begin{align}\label{def:AO2}
 \A(R)		&:= \bigvee_{R\supset\OO\in\OOO}\A(\OO)\,.
\end{align}
This prescription determines in particular the locally generated subalgebras $\A(W_R)\subset\M$, $\A(W_L)\subset\M'$ associated to the right and left wedges, and the von Neumann algebra $\A(\Rl^2)\subset\B(\Hil)$ of all local observables.
\\\\
It is straightforward to verify that the so defined algebras $\A(\OO)$ comply with the basic principles of isotony, locality and covariance \cite{haag}, i.e. they fulfill \cite[Sec. III]{borchers-2d}, $\OO,\OO_1,\OO_2\subset\Rl^2$,
\begin{itemize}
 \item[]{\bf Isotony:}
 \begin{align}
  \A(\OO_1)\subset\A(\OO_2)\qquad{\rm for}\qquad\OO_1\subset\OO_2\,.
 \end{align}
 \item[]{\bf Locality:}
 \begin{align}
  \A(\OO_1)\subset\A(\OO_2)'\qquad{\rm for}\qquad \OO_1\subset\OO_2'\,.
 \end{align}
  \item[]{\bf Translation Covariance:}
  \begin{align}
   U(x)\A(\OO)U(x)^{-1}=\A(\OO+x)\,,\qquad x\in\Rl^2\,.
  \end{align}
\end{itemize}
As a consequence of A2), the modular theory of Tomita and Takesaki (cf., for example, \cite{KadRin}) applies to the pair $(\M,\Om)$. It has been shown by Borchers \cite{borchers-2d} that in the present situation, the modular unitaries and modular group of $(\M,\Om)$ can be used to extend $U$ to a representation of the proper Poincar\'e group, under which the net $\OO\mapsto\A(\OO)$ also transforms covariantly. But this fact will not be needed in our subsequent considerations.
\\\\
The three properties of the algebras $\A(\OO)$ mentioned above allow to interpret the elements of $\A(\OO)$ as observables which are localized in $\OO\subset\Rl^2$. However, two important properties of these algebras are still missing. 

First of all, it is not clear if our definition contains any non-trivial observables localized in bounded spacetime regions, i.e. the intersections \eqref{def:AO} could be trivial in the sense that $\A(\OO)=\Cl\cdot 1$. Since a quantum field theory should contain local observables, at least in spacetime regions above some minimal size, a condition implying the non-triviality of the algebras \eqref{def:AO} is necessary.

Thinking of applications to the explicit construction of models in an inverse scattering approach, one would also like to implement the postulate that the models defined by the observable algebras $\A(\OO)$ have a well-defined S-matrix.

In the framework of algebraic quantum field theory, collision states and the S-matrix can be calculated with the help of Haag-Ruelle scattering theory \cite{araki}. For this method to be applicable, however, two additional conditions have to be satisfied. Firstly, more detailed information about the energy momentum spectrum encoded in $U$ is needed \cite{araki}. As we can choose $U$ from the outset, this requirement poses no difficulties here. But besides these spectral properties, scattering theory relies on the notion of quasi-localized excitations of the vacuum, which can only be constructed if the Reeh-Schlieder property \cite{streater,araki} holds, i.e. if the vacuum vector $\Om$ is cyclic for the local algebras $\A(\OO)$.  
\\
\\
In \cite{BuLe}, the following additional assumption, known as the {\em modular nuclearity condition} in the literature \cite{BDL1,BDL2}, was made to exclude the case of a quantum field theory without local observables.
\begin{itemize}
\item[A4)] Let $\Delta$ denote the modular operator of $(\M,\Om)$. Then for $x\in W_R$, the maps
\begin{align}\label{def:Xi}
 \Xi(x):\M\to\Hil,\qquad \Xi(x)A:=\Delta^{1/4}U(x)A\Om
\end{align}
are assumed to be nuclear\footnote{The definition of a nuclear map between two Banach spaces is recalled in Section \ref{sec:nuc}.}.
\end{itemize}
The modular nuclearity condition A4) is known \cite{BDL1} to imply the split property \cite{dolo} for the inclusion $\M(x)\subset\M$, $x\in W_R$, i.e. the existence of a type I factor $\NN_x$ such that
\begin{align}
 \M(x)\subset\NN_x\subset\M\,.
\end{align}
Moreover, it has been shown in \cite{BuLe} that this inclusion is even a {\em standard split inclusion} in the terminology of \cite{dolo}, i.e. there exist vectors in $\Hil$ which are cyclic and separating for the three algebras $\M$, $\M(x)$ and $\M\cap\M(x)'$. Since $\M$ (and hence $\M(x)$, too) is a factor (Lemma \ref{lem:m31}), the standard split property of $\M(x)\subset\M$ is equivalent to the following condition \cite{dalo,dolo}.
\begin{itemize}
\item[A4$'$)] For $x\in W_R$, there exists a unitary $V_x:\Hil\to\Hil\otimes\Hil$ such that\footnote{We write $\M'\vee\M(x)$ to denote the von Neumann algebra generated by $\M'$ and $\M(x)$.}
\begin{align}
 V_x\big(\M'\vee\M(x)\big)V_x^* &= \M'\otimes\M(x)\,,\label{splituni1}\\
 V_x M' N V_x^*	&=	M'\otimes N,\qquad M'\in\M',\;N\in\M(x)\,.\label{splituni2}
\end{align}
\end{itemize}
Given A1)-A3), condition A4) implies A4$'$), but A4$'$) is slightly weaker than A4) \cite{BDL1}. In the present section, devoted to a model-independent analysis of the algebraic structure, A4$'$) turns out to be the more convenient condition to work with. However, the somewhat stronger modular nuclearity condition A4) has the advantage that it can be checked more easily in concrete applications. We therefore formulate also the results of this section in terms of the latter condition, and begin by recalling the non-triviality result of \cite{BuLe}.
\begin{theorem}{\bf\cite{BuLe}}\label{thm1} Consider a triple $(\M,U,\Hil)$ satisfying the assumptions A1)-A4). Then the double cone algebras $\A(\OO)$ \eqref{def:AO}, $\OO\in\OOO$, are isomorphic to the hyperfinite type III$_1$ factor.
\end{theorem}

As type III algebras, the double cone algebras are far from trivial, and therefore local observables exist in abundance if the modular nuclearity condition is satisfied. Moreover, it follows that the set of vectors which are cyclic for a given $\A(\OO)$, $\OO\in\OOO$, is dense $G_\delta$ in $\Hil$ \cite{BuLe}. In the remainder of this section, we will show that it can also be deduced in the general situation described by the assumptions A1)-A4).

In a particular example of a triple $(\M,U,\Hil)$, the cyclicity of the vacuum has already been established by Buchholz and Summers by explicit calculation of local observables \cite{BuSu-1}. 
\\\\
In the following lemma, we start our analysis by comparing $\M$ to the locally generated wedge algebra $\A(W_R)\subset\M$ \eqref{def:AO2}.
\begin{lemma}\label{lemma:add}
Consider a triple $(\M,U,\Hil)$ satisfying the assumptions A1)-A4). Then $\M$ is locally generated, i.e. $\A(W_R)=\M$.
\end{lemma}
\begin{proof}
Let $x\in W_R$ be fixed and consider the sequence of double cones 
\begin{align}
  \OO_n := \OO_{0,n x}=W_R\cap (W_L+nx)\,,\qquad n\in\N\,.
\end{align}
As $x\in W_R$, this sequence is increasing in the sense that $\OO_n\subset\OO_{n+1}$, $n\in\N$. Moreover, it exhausts all of $W_R$, i.e. every bounded subset of $W_R$ lies in some $\OO_n$. Hence $\bigvee_n\A(\OO_n)=\A(W_R)$. 

The left vertex of each $\OO_n$ is the origin, and the algebras $\A(\OO_n)$ are according to the definition \eqref{def:AO} given by $\A(\OO_n)=\M\cap \M(nx)'$.
 
The split property \eqref{splituni1} provides us with a unitary $V_x:\Hil\to\Hil\otimes\Hil$ implementing an isomorphism between $\A(\OO_1)'=\M'\vee\M(x)$ and $\M'\otimes\M(x)$. In view of \eqref{splituni2}, we find by restriction to $\M'$
\begin{equation}\label{split-restrict}
 V_x\M'V_x^*=\M'\otimes 1\,,
\end{equation}
and since $\M(nx) \subset \M(x)$ (A3), also 
\begin{equation}
 V_x\A(\OO_n)'V_x^* = V_x\big(\M'\vee\M(nx)\big)V_x^*= \M'\otimes\M(nx)\,,\qquad n\in\N\,.
\end{equation}
We thus obtain
\begin{align}
    \A(W_R)'
    &=
    \bigcap_{n\in\mathbb{N}} \A(\OO_n)'
    =
    V_x^* \bigcap_{n\in\mathbb{N}} \Big( \M'\otimes\M(nx)\Big)\,V_x\nonumber
    \\
   &= V_x^* \Big(\M'\otimes\bigcap_{n\in\mathbb{N}}\M(nx)\Big)\,V_x.\label{eq:tensorintersection}
\end{align}
In the last step, we used the commutation theorem for tensor products of von Neumann algebras (or rather, a consequence thereof, see \cite[Cor. IV.5.10]{Takesaki1}).
 
Now consider $\M_\infty:=\bigcap_n\M(nx)$. By construction, this algebra is stable under translations, $U(y)\M_\infty U(y)^{-1} \subset \M_\infty$, $y\in\Rl^2$. The same is true for its commutant $\M_\infty'$, which furthermore has $\Om$ as a cyclic vector since it contains $\M'$, and $\Om$ is cyclic for $\M'$ (A2). But as $\Om$ is (up to multiples) the only translation invariant vector and the spectrum condition holds (A1), it follows by standard arguments (cf., for example, \cite{borchers-halfsided}) that $\M_\infty'=\B(\Hil)$, i.e. $\M_\infty=\Cl\cdot 1$.
 
Inserting this equality into \eqref{eq:tensorintersection} yields $\A(W_R)'=V_x^*(\M'\otimes 1)V_x$, which in view of \eqref{split-restrict} equals $\M'$. Hence the claim $\A(W_R)=\M$ follows.
\end{proof}
To go on, we need another lemma, which is due to M\"uger \cite[Lemma 2.7]{mueger}.
\begin{lemma}[M\"uger]\label{lemma:mueger}
Let $(\M,U,\Hil)$ satisfy the assumptions A1)-A4), and consider three points $x,y,z\in\Rl^2$ such that $y-x\in W_R$ and $z-y\in W_R$.

Then $\A(\OO_{x,y})\vee\A(\OO_{y,z})=\A(\OO_{x,z})$.
\end{lemma}
Geometrically speaking, this lemma states that the algebras of two double cones $\OO_1,\OO_2$ having one of their (left or right) vertices in common generate the algebra of the smallest double cone containing $\OO_1$ and $\OO_2$. It thus establishes a relation between the algebras of double cones of different sizes, and enables us to derive the Reeh-Schlieder property.
 
In the following theorem, we give a proof of this property and some related consequences of Lemma \ref{lemma:add} and \ref{lemma:mueger}.
\begin{theorem}\label{thm2}
Consider a triple $(\M,U,\Hil)$ satisfying A1)-A4) and the local net $\OO\mapsto\A(\OO)$ defined by (\ref{def:AO},\ref{def:AO2}). Then
\begin{enumerate}
\item The Reeh-Schlieder property holds, i.e. $\Om$ is cyclic and separating for each double cone algebra $\A(\OO)$, $\OO\in\OOO$.
\item Haag duality holds, i.e. $\A(\OO)'=\A(\OO')$ for any double cone or wedge $\OO$.
\item Weak additivity holds, i.e. for any open region $\OO\subset\Rl^2$, 
\begin{equation}
 \bigvee_{x\in\Rl^2}\A(\OO+x)=\A(\Rl^2)=\B(\Hil)\,.
\end{equation}
\end{enumerate}
\end{theorem}
\begin{proof}
a) Given any two double cones $\OO,\tilde{\OO}\in\OOO$, there exists $n\in\N$ and translations  $x_1,...,x_n\in\Rl^2$ such that
\begin{align}
 \tilde{\OO} \subset \left( (\OO+x_1)\cup ... \cup(\OO+x_n)\right)''\,,
\end{align}
and $(\OO+x_k)$, $(\OO+x_{k+1})$ have one vertex in common, $k=1,...,n-1$. So, by iterated application of Lemma \ref{lemma:mueger}, it follows that 
\begin{align}
 \bigvee_{x\in\Rl^2}\A(\OO+x) = \bigvee_{\tilde{\OO}\in\OOO}\A(\tilde{\OO}) = \A(\Rl^2)\,.
\end{align}
Hence we can use the standard Reeh-Schlieder argument making use of the spectrum condition of $U$ (cf., for example, \cite{araki}) to show that $\Om$ is cyclic for $\A(\OO)$ if and only if it is cyclic for $\A(\Rl^2)$. But in view of Lemma \ref{lemma:add},  $\M=\A(W_R)$ is contained in $\A(\Rl^2)$. Since $\Om$ is cyclic for $\M$, it is also cyclic for $\A(\Rl^2)$, and hence for $\A(\OO)$.

It has been shown by Borchers \cite{borchers-2d} that the modular conjugation $J$ of $(\M,\Om)$ acts as the total spacetime reflection, $J\A(\OO)J=\A(-\OO)$. Since $W_L=-W_R$, this implies together with Lemma \ref{lemma:add} $\A(W_L)=J\M J=\M'$, i.e. wedge duality holds. Taking into account the translation covariance of the net, we furthermore find
\begin{align}
 \A(\OO_{x,y})'	= \M(x)'\vee\M(y) = \A(W_L+x)\vee\A(W_R+y) = \A({\OO_{x,y}}')\,,
\end{align}
showing the Haag duality of the net (b).

According to the above remarks, $\A(\Rl^2)$ contains $\M$ and $\M'$. Since $\M$ is a factor, also $\A(\Rl^2)=\B(\Hil)$ follows, as the last claim to be proven.
\end{proof}

Besides the properties of the net $\OO\mapsto\A(\OO)$ mentioned in Theorem \ref{thm2}, further additional features like the split property for double cones, the time slice property and $n$-regularity can be derived, as has been shown by M\"uger \cite{mueger}.
\\\\
The strong results of Theorems \ref{thm1} and \ref{thm2} open up a new perspective on the construction of quantum field theories on two-dimensional Minkowski space, emphasizing the role of the local observable algebras. Each triple $(\M,U,\Hil)$ satisfying the assumptions A1)-A4) gives rise to a net of local algebras, which can be interpreted as the (non-trivial) observable algebras of a well-defined quantum field theory satisfying the Reeh-Schlieder property. 

Hence model theories can be constructed by finding examples of such triples. In the following sections, we will construct these objects, verify the assumptions A1)-A4), and discuss the properties of the corresponding model theories.

\section{A Class of Models with Factorizing S-Matrices}\label{sec:model}

We now turn to the concrete construction of interacting quantum field theories on two-dimensional Minkowski space. For simplicity, we consider here models containing only a single species of particles\footnote{The extension of the program to models with a richer particle spectrum is currently under investigation.} of mass $m>0$. We will use the rapidity $\te$ to parametrize the (one-dimensional) upper mass shell according to $p(\te):=m(\cosh\te,\sinh\te)$.
 
Our approach is that of inverse scattering theory, i.e. a given S-matrix $S$ is the input in the construction. The family of theories we will study is characterized by the condition that $S$ is {\em factorizing}. (For an introduction to factorizing S-matrices, see for example the review \cite{dorey}.) This term derives from the fact that in a model with a factorizing S-matrix, all scattering amplitudes are products of delta distributions and a single function \cite{iagol}, the so-called {\em scattering function} $S_2$. On rapidity wavefunctions $\Psi_n^{\rm in}(\te_1,...,\te_n)$ of $n$ incoming particles, the S-matrix $S$ therefore acts as a multiplication operator,
\begin{equation}\label{S-mult}
 (S\Psi_n^{\rm in})(\te_1,...,\te_n)
 =
 \prod_{1\leq l<k\leq n}S_2(|\te_l-\te_k|)
 \cdot
 \Psi_n^{\rm in}(\te_1,...,\te_n)
 \,.
\end{equation}
In particular, the particle number is a conserved quantity in collision processes governed by a factorizing S-matrix. This feature is typical for completely integrable models, which provide a rich class of examples for such scattering operators \cite{abdalla}.

Basic properties of $S$, like unitarity, crossing symmetry and its analytic properties, imply corresponding properties of the scattering function $S_2$ \cite{abdalla,dorey,kb2}, which we take as a definition. Here and in the following, we write
\begin{align}\label{def:strip}
 S(a,b) := \{\zeta\in\Cl\,:\,a<{\rm Im}\,\zeta<b\}
\end{align}
for strips in the complex plane.
\begin{definition}{\bf(Scattering functions)}\label{def:S2}\\
A scattering function is a bounded and continuous function $S_2:\overline{S(0,\pi)}\to\Cl$ which is analytic in the interior of this strip and satisfies, $\te\in\Rl$
\begin{equation}\label{s2rel}
	\overline{S_2(\te)}		=	S_2(\te)^{-1}	= 	S_2(\te+i\pi)		=	S_2(-\te)\,.
\end{equation}
The set of all scattering functions is denoted $\SF$.
\end{definition}

A special scattering function is $S_2^{\rm free}(\te)=1$, which belongs to the interaction-free S-matrix $S^{\rm free}={\rm id}$ \eqref{S-mult}. A simple example with non-trivial interaction is given by the Sinh-Gordon model. By comparison with perturbation theory, the scattering function of this model is expected to be \cite{AFZ}
\begin{equation}
 S_2^{\rm ShG}(\te)
 =
 \frac{\sinh\te-i\sin b}{\sinh\te+i\sin b}\,,
\end{equation}
where the parameter $b$ is related to the coupling constant $g$ of the Sinh-Gordon Lagrangian by $b=\pi g^2(4\pi+g^2)^{-1}$.
\\
\\
Fixing $S_2\in\SF$, we now recall the construction of an associated triple $(\M,U,\Hil)_{S_2}$ consisting of a Hilbert space $\Hil$, a representation $U$ of the translations on $\Hil$, and a ``wedge algebra'' $\M\subset\B(\Hil)$ of the type studied in Section \ref{sec:alg}. For details, we refer the reader to \cite{gl1,gl-phd,S1}.

To describe the Hilbert space, we introduce on $L^2(\Rl^n)$ an $S_2$-dependent representation $D_n$ of the group $\frS_n$ of permutations of $n$ letters. Given $\rho\in\frS_n$, we put
\begin{align}\label{def:Dn}
\left(D_n(\rho) f_n\right)(\te_1,...,\te_n)
	&=
	S^\rho(\te_1,...,\te_n)\cdot f_n(\te_{\rho(1)},...,\te_{\rho(n)}), 
	\\
	S^\rho(\te_1,...,\te_n)
	&:=
	\prod_{\genfrac{}{}{0pt}{}{1\leq l < k \leq n}{\rho(l) > \rho(k)}} S_2(\te_{\rho(l)}-\te_{\rho(k)})\,.
\end{align}
In particular, the transpositions $\tau_j$, $j=1,...,n-1$, are represented as
\begin{equation}\label{dn-trans}
 (D_n(\tau_j)f_n)(\te_1,...,\te_n)	=	S_2(\te_{j+1}-\te_j)\cdot f_n(\te_1,...,\te_{j+1},\te_j,...,\te_n)\,.
\end{equation}
Using the properties \eqref{s2rel} of $S_2$, it has been shown in \cite{gl-phd} (see also \cite{LiMi,gl1}) that $D_n$ is a unitary representation of $\frS_n$ on $L^2(\Rl^n)$, and that the mean over $D_n$,
\begin{equation}\label{def:Pn}
 P_n := \frac{1}{n!}\sum_{\rho\in\frS_n}D_n(\rho)\,,
\end{equation}
is the orthogonal projection onto the $D_n$-invariant functions in $L^2(\Rl^n)$.

With the help of the ``$S_2$-symmetrization'' $P_n$, we define the Hilbert space $\Hil$ of the model with scattering function $S_2$ as
\begin{align}\label{def:Hil}
 \Hil_0	&:=	\Cl\,,\qquad \Hil_n:=P_nL^2(\Rl^n),\quad n\geq 1\,,\qquad \Hil:=\bigoplus_{n=0}^\infty\Hil_n\,.
\end{align}
The vectors in $\Hil$ are sequences $\Psi=(\Psi_0,\Psi_1,...\,)$, $\Psi_n\in\Hil_n$, such that the norm corresponding to the scalar product $\langle\Psi,\Phi\rangle:=\overline{\Psi_0}\Phi_0+\sum_{n=1}^\infty\langle\Psi_n,\Phi_n\rangle_{L^2(\Rl^n)}$ is finite. Due to its invariance under $D_n(\tau_j)$ \eqref{dn-trans}, a function $\Psi_n\in\Hil_n$ has the symmetry property
\begin{align}\label{eq:S2sym}
	\Psi_n(\te_1,...,\te_{j+1},\te_j,...\te_n) 
  		&=
	S_2(\te_j-\te_{j+1}) \cdot \Psi_n(\te_1,...,\te_j,\te_{j+1},...,\te_n)\,.
\end{align}
Note that for $S_2=1$, this construction yields precisely the Bose Fock space over $\Hil_1$. For generic $S_2\in\SF$, we refer to $\Hil$ as the {\em $S_2$-symmetric Fock space}. 

As a domain for some unbounded operators on $\Hil$, we also introduce the dense subspace of terminating sequences $(\Psi_0,\Psi_1,...,\Psi_n,0,0,...\,)$, $\Psi_k\in\Hil_k$, which will be denoted $\DD$. For example, the particle number operator $N$, $(N\Psi)_n:=n\cdot\Psi_n$, is well defined on $\DD$.

On $\Hil$, there acts a representation $U$ of the proper Poincar\'e group\footnote{Actually, this representation extends to the whole Poincar\'e group $\PG$, but this fact will not be needed here.} $\PG_+$. The proper orthochronous Poincar\'e transformations $(x,\la)\in\PGpo$ consisting of a boost with rapidity parameter $\la\in\Rl$ and a subsequent translation along $x\in\Rl^2$ are represented as, $\Psi\in\Hil$,
\begin{align}\label{def:U}
	\big(U(x,\la) \Psi\big)_n(\te_1,...,\te_n) 
		&:=
	\exp\bigg(i\sum_{k=1}^n p(\te_k) \cdot x \bigg) \cdot \Psi_n(\te_1-\la,...,\te_n-\la)\,,
\end{align}
and the reflection $j(x):=-x$ as
\begin{align}\label{def:Uj}
 (U(j)\Psi)_n(\te_1,...,\te_n)	&:=	\overline{\Psi_n(\te_n,...,\te_1)}\,.
\end{align}
For the proof that $U$ is an (anti-) unitary representation of $\PG_+$ on $\Hil$, see \cite{gl1}. By inspection of $U$, it follows that $\Om:=(1,0,0,...\,)$ is up to multiples the only $U$-invariant vector in $\Hil$.

The representation of the translations required in the discussion of the previous section will be identified with the restriction of $U$ to the translation subgroup, and we will also employ the shorthand notation $U(x):=U(x,0)$. Clearly, $U$ satisfies the spectrum condition and is strongly continuous, i.e. assumption A1) of Section \ref{sec:alg} is satisfied.
\\
\\
Having specified the Hilbert space $\Hil$ and the representation $U$, we now turn to the construction of the ``wedge algebra'' $\M\subset\B(\Hil)$. 

On the $S_2$-symmetric Fock space, there acts an algebra of creation and annihilation operators $\zd(\psi)$, $z(\psi)$, $\psi\in\Hil_1$. These are unbounded operators, in general, but always contain $\DD$ in their domains. They are defined by, $\Phi\in\DD$,
\begin{align}\label{def:zzd}
 (\zd(\psi)\Phi)_n
 :=
 \sqrt{n}\,P_n(\psi\otimes\Phi_{n-1})
 \,,\qquad
 z(\psi)
 :=
 \zd(\overline{\psi})^*
 \,,
 \qquad
 \psi\in\Hil_1\,.
\end{align}
Since $z(\psi)\Om=0$, $\zd(\psi)\Om=\psi$, we call $z$ and $\zd$ annihilation and creation operators, respectively. The following bounds with respect to the particle number operator $N$ hold \cite{BuLe}:
\begin{align}\label{Z-bounds}
	\|z(\psi)\Phi\| 	\leq	\|\psi\|\|N^{1/2}\Phi\|
	\,,\quad
	\|\zd(\psi)\Phi\| 	\leq	\|\psi\|\|(N+1)^{1/2}\Phi\|
	\,,\quad \Phi\in\DD\,.
\end{align}
From time to time, we will also work with the distributions $z(\te)$, $\zd(\te)$, which are related to the above operators by the formal integrals $z^\#(\psi)=\int d\te\,\psi(\te)z^\#(\te)$, $z^\#=z,\zd$. These distributions satisfy the relations of the Zamolodchikov-Faddeev algebra \cite{ZZ,smirnov},
\begin{subequations}\label{ZFdis}
\begin{align}
	z(\te_1)z(\te_2)	&=	S_2(\te_1-\te_2)\,z(\te_2)z(\te_1),\\
	z(\te_1)\zd(\te_2) 	&=	S_2(\te_2-\te_1)\,\zd(\te_2) z(\te_1) + \delta(\te_1-\te_2)\cdot 1 \,,
\end{align}
\end{subequations}
where $1$ denotes the identity in $\B(\Hil)$. For later reference, we mention here that in view of the definition \eqref{def:zzd}, there holds in particular
\begin{equation}\label{eq:zdn}
 \langle\zd(\te_1)\cdots\zd(\te_n)\Om,\,\Psi\rangle
 =
 \sqrt{n!}\,\Psi_n(\te_1,...,\te_n)
 \,,\qquad
 \Psi\in\Hil\,.
\end{equation}

Following Schroer \cite{S1,S2,SW}, the Zamolodchikov operators can be combined to define a quantum field $\phi$. For Schwartz test functions $f\in\Ss(\Rl^2)$, we put
\begin{align}
    \phi(f) := \zd(f^+)+z(f^-)
    \,,\qquad
    f^\pm(\te) := \frac{1}{2\pi}\int d^2x\,f(x)e^{\pm ip(\te)\cdot x}\,.
    \label{def:phi}
\end{align}
This field has a number of interesting properties \cite{gl1}. To begin with, it transforms covariantly under the adjoint action of the proper orthochronous Poincar\'e transformations $U(x,\la)$, and it has the Reeh-Schlieder property. Moreover, $\phi$ is a solution of the Klein-Gordon equation since it creates single particle states from the vacuum. Convenient mathematical properties of $\phi(f)$ are its essential self-adjointness for real $f$, and the fact that $f\mapsto\phi(f)\Psi$, $\Psi\in\DD$, is a vector valued tempered distribution.

As is familiar from free field theory, $\phi$ has well-defined time zero fields $\varphi$, $\pi$, which are given by
\begin{subequations}\label{def:phipi}
 \begin{align}
\varphi(f) &= \zd(\fhat) + z(\fhat_-),
    &\fhat(\te)&:=\fti(m\sinh\te),\\
    \pi(f)     &= i\big(\zd(\omega\fhat)-z(\omega\fhat_-)\big),
    &\fhat_-(\te)&:=\fhat(-\te)\,.
\end{align}
\end{subequations}
Here the single particle Hamiltonian $\omega=m\cosh\te$ acts as a multiplication operator on its domain in $\Hil_1$.

However, as a consequence of the commutation relations \eqref{ZFdis}, $\phi$ is {\em not local}, in general. Locality holds if and only if $S_2=1$, and in this case $\phi$ coincides with the free scalar field of mass $m$.

For generic scattering function $S_2\in\SF$, it was discovered by Schroer \cite{S1} that although $\phi$ is not strictly local, it is not completely delocalized either. The localization properties of $\phi$ are most easily understood by introducing a second field operator $\phi'$ \cite{gl1},
\begin{equation}
\phi'(f) := U(j)\phi(f^j)U(j)\,,\qquad f^j(x):=\overline{f(-x)}\,.
\end{equation}
The two fields $\phi,\phi'$ are relatively wedge-local in the following sense: For (real) test functions $f,g$ with $\supp f\subset W_L$, $\supp g\subset W_R$, the selfadjoint closures of $\phi(f)$ an $\phi'(g)$ commute, i.e. $[e^{i\phi(f)},e^{i\phi'(g)}]=0$. Therefore $\phi(x)$ can be consistently interpreted as being localized in the left wedge $W_L+x$, and $\phi'(y)$ is localized in the right wedge $W_R+y$.

Switching to the algebraic formulation, we consider the ``wedge algebra''
\begin{equation}\label{def:M}
 \M	:=	\big\{e^{i\phi(f)}\,:\,f\in \Ss_\Rl(W_L) \big\}'
 =	\big\{e^{i\phi'(f)}\,:\,f\in \Ss_\Rl(W_R) \big\}''
 \,.
\end{equation}
Here the first equality defines $\M$, and the second is a result of \cite{BuLe}.

For $f\in\Ss(\Rl^2)$ with $\supp f\subset W_L$, the field operator $\phi(f)$ satisfies \cite{gl1}
\begin{align}
 \langle\Psi,\,[\phi(f),A]\,\Phi\rangle=0\,,\qquad A\in\M,\;\Psi,\Phi\in\DD\,.
\end{align}
Analogously, on can show that for testfunctions $h\in\Ss(\Rl)$, $\supp h\subset \Rl_-$,
\begin{align}\label{phipiA}
 \langle\Psi,\,[\varphi(h),A]\,\Phi\rangle=0\,,\quad\langle\Psi,\,[\pi(h),A]\,\Phi\rangle=0\qquad A\in\M,\;\Psi,\Phi\in\DD\,.
\end{align}
The definition of $\M$ completes the data $(\M,U,\Hil)_{S_2}$. The assumptions A2) and A3), regarding the cyclicity and separating property of $\Om$ for $\M$, and the isotony of this algebra under translations inside $W_R$, can be deduced from the Reeh-Schlieder property and the translation covariance of $\phi$ and $\phi'$ \cite{gl1,gl-phd}. We note down these facts as a theorem.
\begin{theorem}{\bf \cite{gl1}}
 Let $S_2\in\SF$. Then the triple $(\M,U,\Hil)_{S_2}$ defined through \eqref{def:Hil}, \eqref{def:U} and \eqref{def:M} satisfies the assumptions A1)-A3) of Section \ref{sec:alg}. 
\end{theorem}
With A1)-A3) fulfilled, we can apply the construction of Section \ref{sec:alg} to define a net $\OO\mapsto\A(\OO)$ of local observable algebras on $\Rl^2$. As discussed there, the crucial question in this context is whether also the modular nuclearity condition A4) holds. If it holds, the existence of observables, the Reeh-Schlieder property, and, as we shall see in Section \ref{sect:s}, the anticipated form \eqref{S-mult} of the S-matrix follow.

If, on the other hand, condition A4) fails, the status of all these important properties is unclear. It is thus doubtful if the inverse scattering problem has a solution, i.e. if there exists a local quantum theory with the considered S-matrix, in that case.

It has been shown in \cite{BuLe} and \cite{GL-1} that the modular nuclearity condition is satisfied for the constant scattering functions $S_2=1$ and $S_2=-1$, respectively. In the following, we will analyze this condition for the class of models with regular scattering functions, defined below.
\begin{definition}{\bf(Regular scattering functions)}\label{def:S0}\\
A scattering function $S_2\in\SF$ is called regular if there exists $\kappa>0$ such that $S_2$ continues to a bounded analytic function on the strip $S(-\kappa,\pi+\kappa)$. Denoting $\tilde{\kappa}$ the maximal value of $\kappa$ compatible with this condition, we define $\kappa(S_2):=\min\{\frac{\pi}{2},\tilde{\kappa}\}$ and
\begin{align}\label{def:S2kappa}
\|S_2\|
&:=
\sup\big\{|S_2(\zeta)|\,:\,\zeta\in \overline{S(-\kappa(S_2),\pi+\kappa(S_2))}\big\}<\infty
\,.
\end{align}
The family of all regular scattering functions is denoted $\SF_0$.
\end{definition}
The two regularity assumptions made in this definition can be understood as follows. As a consequence of the relations $S_2(-\te)=S_2(\te)^{-1}=S_2(\te+i\pi)$ \eqref{s2rel}, $S_2$ can be continued to a meromorphic function on all of $\Cl$. A pole $\zeta$ in the ``unphysical sheet'' $-\pi<{\rm Im}\,\zeta<0$ is usually interpreted as evidence for an unstable particle with a finite lifetime \cite{eden}, and the lifetime of such a resonance becomes arbitrarily long if the corresponding pole lies sufficiently close to the real axis. A scattering function with a sequence of poles in $S(-\pi,0)$ which approach the real axis might therefore have infinitely many almost stable resonances with ``masses'' such that the thermodynamical partition function diverges. 

But the modular nuclearity condition is closely related \cite{BDL2} to the thermodynamically motivated {\em energy nuclearity condition} of Buchholz and Wichmann \cite{BuWi}, and the latter condition might well be violated for the previously described distribution of resonances. We therefore expect the modular nuclearity condition to {\em fail} in this situation (although there might still exist local observables). To exclude such models, we require all singularities of a regular scattering function $S_2$ to lie a finite distance off the real axis, i.e. $S_2$ to continue analytically to a strip of the form $S(-\kappa,\pi+\kappa)$, $\kappa>0$.

The second requirement, postulating $S_2$ to stay bounded also on the enlarged strip, amounts to a condition on the phase shift of $S_2$ (cf. \cite{ktt} for a similar assumption).
\\\\
All scattering functions known from Lagrangian models, like the Sinh-Gordon model, satisfy the regularity assumptions of Definition \ref{def:S0}. Particular examples for $S_2\in\SF_0$ are
\begin{equation}\label{ex:s2}
S_2(\te) = \pm\prod_{k=1}^N \frac{\sinh\beta_k-\sinh\te}{\sinh\beta_k+\sinh\te}
\,,
\qquad
0<{\rm Im}\,\beta_1,...,{\rm Im}\,\beta_N<\pi\,,
\end{equation}
where with each $\beta_k$, also $-\overline{\beta_k}$ is required to be included in the set  $\{\beta_1,...,\beta_N\}$. 
\\
\\
In the following, it is our aim to prove the modular nuclearity condition A4) for the models with scattering functions $S_2\in\SF_0$, i.e. to show that the maps 
\begin{equation}\label{def:XXi}
 \Xi(x):\M\to\Hil\,,\qquad \Xi(x)A:=\Delta^{1/4}U(x)A\Om\,,\qquad x\in W_R,
\end{equation}
are nuclear maps between the Banach spaces $(\M,\|\cdot\|_{\B(\Hil)})$ and $(\Hil,\|\cdot\|)$.

This task is facilitated by the fact that in the models at hand, the modular data $J,\Delta$ of $(\M,\Om)$ are known to act geometrically ``correct'', i.e. as expected from the Bisognano-Wichmann theorem \cite{BiWi1,BiWi2}. More precisely, the modular conjugation coincides with the TCP operator, $J=U(j)$ \eqref{def:Uj}, and the modular unitaries are given by the boost transformations $\Delta^{it}=U(0,-2\pi t)$ \cite{BuLe}. In particular, $\Delta^{1/4}U(x)$ commutes with the projection $P_n$ onto $\Hil_n$ \eqref{def:Pn}, such that the $n$-particle restrictions $\Xi_n(x)$ of $\Xi(x)$ take the form
\begin{equation}\label{def:Xin}
 \Xi_n(x):\M\to\Hil_n\,,\qquad \Xi_n(x)A:=P_n\Xi(x)A=\Delta^{1/4}U(x)(A\Om)_n\,.
\end{equation}
We now consider a purely spatial translation $x=(0,s)=:\us$, $s>0$, and write $\Xi_n(s)$ instead of $\Xi_n(\us)$. The notation $\us=(0,s)$ will be used throughout the following sections, and the parameter $s>0$ will be referred to as the {\em splitting distance}.

As $\Delta^{1/4}$ acts as the boost with imaginary rapidity parameter $\frac{i\pi}{2}$, and since $i\,p(\te-\frac{i\pi}{2})\cdot (0,s)=-ms\cosh\te$, the maps $\Xi_n(s)$ are explicitly given by, $A\in\M$,
\begin{equation}\label{def:xins}
 \left(\Xi_n(s)A\right)_n(\te_1,...,\te_n)
 =
 \prod_{k=1}^n e^{-ms\cosh\te_k}\cdot (A\Om)_n(\te_1-\tfrac{i\pi}{2},...,\te_n-\tfrac{i\pi}{2})\,.
\end{equation}
The right hand side has to be understood in terms of analytic continuation, and suggests to study the analytic properties of $(A\Om)_n$ for the proof of the nuclearity condition. This is done in the subsequent Section \ref{sec:ana}. In Section \ref{sec:nuc}, the nuclearity of the maps \eqref{def:xins} will then be established.

\section{Analytic Properties of Wedge-Local Form Factors}\label{sec:ana}
In this section, we consider a regular scattering function $S_2\in\SF_0$ and a fixed operator $A$ in the associated wedge algebra $\M$ \eqref{def:M}. We will study analyticity and boundedness properties of the $n$-particle rapidity functions $(A\Om)_n = P_nA\Om$. These functions are precisely the form factors of $A$ \eqref{eq:zdn},
\begin{equation}\label{aff}
 (A\Om)_n(\te_1,...,\te_n)=\frac{1}{\sqrt{n!}}\langle\zd(\te_1)\cdots\zd(\te_n)\Om,A\Om\rangle\,.
\end{equation}

Our notation will be as follows. Vectors in $\Rl^n$ and $\Cl^n$ are denoted by boldface letters $\bla, \bte, \bze$, and their components by $\la_k,\te_k,\zeta_k$. As multidimensional generalizations of the strip regions $S(a,b)$ \eqref{def:strip}, we will consider tubes of the form $\Tu:=\Rl^n+i\,\Cu\subset\Cl^n$, where the base $\Cu$ is an open  convex domain in $\Rl^n$. For functions $F:\Tu\to\Cl$, we introduce the notation $F_\sbla(\bte):=F(\bte+i\bla)$, $\bla\in\Cu$.

The main result of the present section is Proposition \ref{prop:ana}, stating that $(A\Om)_n$ is the boundary value of a function analytic in some tube in $\Cl^n$, and that this function is bounded on $\Rl^n+i\,\Cu$, where $\Cu\subset\Rl^n$ is a neighborhood of the point $(-\frac{\pi}{2},...,-\frac{\pi}{2})$ corresponding to the action of the modular operator in \eqref{def:xins}.
\begin{lemma}\label{lemma:chat}
Let $A\in\M$, $n_1,n_2\in\N_0$, $\Psi_{n_1}\in\Hil_{n_1}$, $\Phi_{n_2} \in \Hil_{n_2}$. There exists a function $K:\overline{S(-\pi,0)}\to\Cl$ which is analytic in the interior of this strip and whose boundary values satisfy, $\te\in\Rl$,
\begin{align}\label{c-cd}
   K(\te) = \langle\Psi_{n_1},\,[z(\te),A]\,\Phi_{n_2}\rangle,
   \qquad
   K(\te-i\pi) = -\langle\Psi_{n_1},\,[\zd(\te),A]\,\Phi_{n_2}\rangle \,,
\end{align}
in the sense of distributions. Moreover, with $c(n_1,n_2) := \sqrt{n_1+1}+\sqrt{n_2+1}$, there holds the bound
\begin{align}\label{Chardybound}
\left(\int d\te\, |K(\te-i\la)|^2\right)^{1/2} \leq  c(n_1,n_2)\|\Psi_{n_1}\|\|\Phi_{n_2}\|\|A\|
\,,\qquad 0\leq\la\leq\pi.
\end{align}
\end{lemma}
\begin{proof}
Consider the distributions $K^\# : \Ss(\Rl)\to\Cl$, 
\begin{align}
 K^\#(\fhat) := \langle \Psi_{n_1},[z^\#(\fhat),A]\,\Phi_{n_2}\rangle\,,\qquad z^\#=z,\zd\,.
\end{align}
In view of the bounds \eqref{Z-bounds}, there holds
\begin{align}
|K(\fhat)| &\leq \|\zd(\overline{\fhat})\Psi_{n_1}\|\|A\Phi_{n_2}\| + \|A^*\Psi_{n_1}\|\|z(\fhat)\Phi_{n_2}\|
\\
&\leq \left(\sqrt{n_1+1}+\sqrt{n_2}\right) \|\Psi_{n_1}\|\|\Phi_{n_2}\|\|A\|\cdot\|\fhat\|\,,
\\
|K^\dagger(\fhat)| 
&\leq \left(\sqrt{n_1}+\sqrt{n_2+1}\right) \|\Psi_{n_1}\|\|\Phi_{n_2}\|\|A\|\cdot\|\fhat\|\,.
\end{align}
By application of Riesz' Lemma, it follows that both distributions, $K$ and $K^\dagger$, are given by integration against functions in $L^2(\Rl)$ (denoted by the same symbols) with norms
\begin{align}
 \|K^\#\|_2\leq c(n_1,n_2)\|\Psi_{n_1}\|\|\Phi_{n_2}\|\|A\|\,.\label{bnd-Kzaun}
\end{align}
To obtain the analytic continuation of $K$, we consider the time zero fields $\varphi$, $\pi$ of $\phi$ \eqref{def:phipi}, and the corresponding expectation values $k_\pm:\Ss(\Rl)\to\Cl$, 
 \begin{align}\label{cpm}
  k_-(f) &:= \langle\Psi_{n_1},\,[\varphi(f),A]\,\Phi_{n_2}\rangle\,,
  \qquad
  k_+(f) := \langle\Psi_{n_1},\,[\pi(f),A]\,\Phi_{n_2}\rangle\,.
\end{align}
As a consequence of the localization of $\phi$ in the left wedge and $A$ in the right wedge, $k_\pm(f)$ vanishes for test functions with support on the left half line $\Rl_-$ \eqref{phipiA}. Hence the Fourier transforms of $k_\pm$ are the boundary values of functions $p\mapsto\tilde{k}_\pm(p)$ analytic in the lower half plane, which satisfy polynomial bounds at the boundary and at infinity \cite[Thm. IX.16]{SimonReed2}.

Since $\sinh(\cdot)$ maps $S(-\pi,0)$ to the lower half plane, the functions
\begin{align}\label{def:chat}
  K_+(\te) := \tilde{k}_+(m\sinh\te)
  \,,\qquad
  K_-(\te) := m\cosh\te\cdot\tilde{k}_-(m\sinh\te),
\end{align}
are analytic in the strip $S(-\pi,0)$. The relation between $K_\pm$ and $K^\#$ is found by expressing $z^\#$ in terms of the time zero fields $\varphi$, $\pi$ of $\phi$ \eqref{def:phipi},
\begin{eqnarray}
 \zd(\fhat)	&=&	\frac{1}{2}(\varphi(f)-i\pi(\omega^{-1}f))\,,\qquad \fhat(\te)=\fti(m\sinh\te)\,,\\
 z(\fhat)	&=&	\frac{1}{2}(\varphi(f_-)+i\pi(\omega^{-1}f_-))\,,\qquad f_-(x)=f(-x)\,.
\end{eqnarray}
For the annihilation operator this yields, $f\in\Ss(\Rl)$,
\begin{align*}
  \int d\te\,
  K(\te)\fhat(\te)&= \langle\Psi_{n_1},[z(\fhat),A]\Phi_{n_2}\rangle
  =
  \frac{1}{2}(k_-(f_-)+i\,k_+(\omega^{-1}f_-))\\
  & =
  \frac{1}{2}\int dp\,
  \bigg(\tilde{k}_-(p)+\frac{i\,\tilde{k}_+(p)}{\sqrt{p^2+m^2}}\bigg)\fti(p)\\
  &=
  \frac{1}{2}\int d\te\,\big(K_-(\te)+iK_+(\te)\big)\fhat(\te)
  \,.
\end{align*}
Similarly, one obtains for the creation operator
\begin{align*}
  \int d\te\,
  K^\dagger(\te)\fhat(\te) &=  \frac{1}{2}\int d\te\,\big(K_-(-\te)-iK_+(-\te)\big)\fhat(\te)
  \,.
\end{align*}
It follows from these equations that the boundary values of $K_\pm$ exist as square integrable {\em functions}, and we have the identities
\begin{align}
 K(\te)=\frac{1}{2}\big(K_-(\te)+iK_+(\te)\big)\,,\qquad  K^\dagger(\te)=\frac{1}{2}\big(K_-(-\te)-iK_+(-\te)\big)\,.
\end{align}
Hence $K$ is analytic in $S(0,\pi)$, too, and since $K_\pm(\te-i\pi)=\pm K_\pm(-\te)$ holds for $\te\in\Rl$ \eqref{def:chat}, also the claimed relation $K(\te-i\pi)=-K^\dagger(\te)$ \eqref{c-cd} follows.

It remains to prove the $L^2$-bound \eqref{Chardybound}. Consider the ``shifted'' function $K^{(s)}(\zeta):=e^{-ims\,\sinh\zeta}\cdot K(\zeta)$, $s>0$,
\begin{align}
  \big|K^{(s)}_{-\la}(\te)\big|
  &=
  \frac{1}{2}\, e^{-ms\sin\la\, \cosh\te}
  \,\big|K_-(\te-i\la)+iK_+(\te-i\la)\big|
  \;.\label{mono}
\end{align}
As $\te\mapsto K_\pm(\te-i\la)$ are bounded by polynomials in $\cosh\te$ for $|\te|\to\infty$, $0<\la<\pi$, we have $K^{(s)}_{-\la}\in
L^2(\Rl)$ for all $\la\in[0,\pi]$, $s>0$. In view of the previous estimates \eqref{bnd-Kzaun} on the $L^2$-norms of the boundary values of $K$, the three lines theorem can be applied and we conclude
\begin{align}\label{l2bnd}
  \|K^{(s)}_{-\la}\|_2
  \leq
   c(n_1,n_2)\|\Psi_{n_1}\|\|\Phi_{n_2}\|\|A\|
  \;,\qquad
  0\leq \la\leq\pi
  \;.
\end{align}
But as \eqref{mono} is monotonically increasing as $s\to 0$, this uniform bound holds also for $K_{-\la}=K_{-\la}^{(0)}$, $0\leq\la\leq\pi$.
\end{proof}
Lemma \ref{lemma:chat} is our basic tool for deriving analytic properties of the functions $(A\Om)_n$. In the following, we study matrix elements of the form 
\begin{align*}
\langle \zd(\te_{k+1})\cdots\zd(\te_n)\Om,\,A\,\zd(\te_k)\cdots\zd(\te_1)\Om\rangle\,,
\end{align*}
with certain contractions between the rapidity variables $\te_{k+1},...,\te_n$ in the left and $\te_1,...,\te_k$ in the right argument of the scalar product.

Some notation needs to be introduced. Given two integers $0\leq k\leq n$, we define a {\em contraction} $C$ to be a set of pairs, $C=\{(l_1,r_1),...,(l_N,r_N)\}$, with pairwise different ``left indices'' $l_1,...,l_N\in\{k+1,...,n\}$ and ``right indices'' $r_1,...,r_N\in\{1,...,k\}$. The set of all such contractions is denoted $\CC_{n,k}$. The number $N$ of pairs $(l_j,r_j)$ in a given $C\in\CC_{n,k}$ will be called the length of $C$, and notated as $|C|:=N\leq\min\{k,n-k\}$. We also write $\bl_C:=\{l_1,...,l_N\}$ and $\br_C:=\{r_1,...,r_N\}$ for the sets of left and right indices of the contraction $C$. 

With these notations, a contracted matrix element of $A$ is defined as
\begin{align}\label{def-arl}
\langle \bl_C |\,A\,|\br_C\rangle_{n,k}
&:=
\langle
\zd_{k+1}\cdots\widehat{\zd_{l_1}}\cdots\widehat{\zd_{l_{|C|}}}\cdot\cdot\,\zd_n\Om
\,,A\,
\zd_k\cdots\widehat{\zd_{r_1}}\cdots\widehat{\zd_{r_{|C|}}}\cdot\cdot\zd_1\Om
\rangle\,,
\end{align}
where $\zd_a := \zd(\te_a)$ is considered as an operator-valued distribution in $\te_a$ and the hats indicate omission of the corresponding creation operators. Given the particle number bounds \eqref{Z-bounds} and the boundedness of $A$, we can apply the nuclear theorem to conclude that these contracted matrix elements are well-defined tempered distributions on $\Ss(\Rl^{n-2|C|})$.

For square-integrable functions $F_L\in L^2(\Rl^{n-k-|C|})$ and $F_R\in L^2(\Rl^{k-|C|})$ depending on $\{\te_{k+1},...,\te_n\}\backslash\{\te_{l_1},...,\te_{l_{|C|}}\}$ and $\{\te_1,...,\te_k\}\backslash\{\te_{r_1},...,\te_{r_{|C|}}\}$, respectively, there hold the bounds (cf. \eqref{eq:zdn})
\begin{align}\label{lr-bnd}
\!\!\!\left|\langle\bl_C|\,A\,|\br_C\rangle_{n,k}(F_L\otimes F_R)\right|
&\leq
\sqrt{(n-k-|C|)!}\sqrt{(k-|C|)!}\,\|F_L\| \|F_R\|\|A\|\,.
\end{align}

Employing the shorthand notations $\delta_{l,r}:=\delta(\te_l-\te_r)$ and 
\begin{align}
S_{a,b}		:=	S_2(\te_a-\te_b),\qquad
S^{(k)}_{a,b}
			:=
			\left\{
			\begin{array}{llll}
			S_{b,a}	&\;;\;&	& a\leq k<b\;\;{\rm or}\;\; b\leq k< a\\
			S_{a,b}	&\;;\;&	& {\rm otherwise}
			\end{array}
			\right.
\,,
\end{align}
we associate with each contraction $C=\{(l_1,r_1),...,(l_{|C|},r_{|C|})\}\in\CC_{n,k}$ the following distribution $\delta_C$ and function $S_C^{(k)}$:
\begin{align}\label{def-dS}
\delta_C		&:=	(-1)^{|C|}\prod_{j=1}^{|C|} \delta_{l_j,r_j}
\,,\qquad 
S_C^{(k)}		:=	\prod_{j=1}^{|C|} \prod_{m_j=r_j+1}^{l_j-1}
				S^{(k)}_{m_j,r_j}
				\cdot
				\prod_{\genfrac{}{}{0pt}{}{r_i<r_j}{l_i < l_j}}
				S^{(k)}_{r_j,l_i}
				\,.
\end{align}
In the following, the main objects of interest are the {\em completely contracted matrix elements of $A$}, defined as 
\begin{align}\label{def-Acon}
\langle A\rangle^{\rm con}_{n,k}
&:=
\sum_{C\in\CC_{n,k}} \delta_C\cdot S_C^{(k)}\cdot\langle\bl_C|\,A\,|\br_C\rangle_{n,k}
\,.
\end{align}
The product $\delta_C\cdot S_C^{(k)}\cdot\langle\bl_C|\,A\,|\br_C\rangle_{n,k}$ is defined in the sense of distributions. Note that the product of $\delta_C$ and $\langle\bl_C|\,A\,|\br_C\rangle_{n,k}$ is well-defined because these distributions act on different variables. Since $S_2\in\SF_0$ can be continued to a bounded analytic function on a strip containing the real axis (cf. Definition \ref{def:S0}), the functions $S_C^{(k)}$ are smooth, and all their derivatives are bounded on $\Rl^n$. Hence \eqref{def-Acon} exists as a tempered distribution on $\Ss(\Rl^n)$. 

To discuss the analytic properties of $\langle A\rangle^{\rm con}_{n,k}$, it is convenient to represent this distribution by two alternative formulae, stated below. 
\begin{lemma}\label{lemma:tech}
Let $\hat{\CC}_{n,k}\subset\CC_{n,k}$ denote the subset of those contractions $C\in\CC_{n,k}$ which do not contract $k+1$, i.e. fulfill $k+1\notin\bl_C$. Then
\begin{align}\label{acon-comm1}
\langle A\rangle^{\rm con}_{n,k}
&=
\sum_{C\in\hat{\CC}_{n,k}}
\delta_C S_C^{(k)}
\langle \bl_C\cup\{k+1\}|\,[z_{k+1},A]\,|\br_C\rangle_{n,k}\,,\\
\langle A\rangle^{\rm con}_{n,k+1}
&=
\sum_{C\in\hat{\CC}_{n,k}}
\delta_C S_C^{(k+1)}
\langle \bl_C\cup\{k+1\}|\,[A,\zd_{k+1}]\,|\br_C\rangle_{n,k}\,.\label{acon-comm2}
\end{align}
\end{lemma}
The proof of Lemma \ref{lemma:tech} is based on the exchange relations of the Zamolodchikov-Faddeev algebra \eqref{ZFdis}; it can be found in the appendix.

The analyticity and boundedness properties of the contracted matrix elements $\langle A\rangle^{\rm con}_{n,k}$ are explained in the following lemma.
\begin{lemma}\label{lemma:recursion}
In a model with scattering function $S_2\in\SF_0$, let $A\in\M$. 
\begin{enumerate}
\item $\langle A\rangle_{n,k}^{\rm con}$ has an analytic continuation in the variable $\te_{k+1}$ to the strip $S(-\pi,0)$, $k\leq n-1$. Its distributional boundary value at Im$\,\te_{k+1}=-\pi$ is given by
\begin{align}
\langle A\rangle_{n,k}^{\rm con}(\te_1,...,\te_{k+1}-i\pi,...,\te_n)
&=
\langle A\rangle_{n,k+1}^{\rm con}(\te_1,...,\te_{k+1},...,\te_n)
\,.
\end{align}
\item There holds the bound, $f_1,...,f_n\in\Ss(\Rl)$, $0\leq\la\leq\pi$,
\begin{equation}\label{acon-bnd}
\!\!\!\!\!\!\left|\int d^n\bte \langle A\rangle_{n,k}^{\rm con}(\te_1,..,\te_{k+1}-i\la,..,\te_n) \prod_{j=1}^n f_j(\te_j)\right|
\leq
2^n\sqrt{n!}\,\|A\|\prod_{j=1}^n\|f_j\|_2
\,.
\end{equation}
\end{enumerate}
\end{lemma}
\begin{proof}
a) Consider the distribution $\langle A\rangle^{\rm con}_{n,k}$, expressed as in \eqref{acon-comm1}. As $k+1$ is not contracted in $C\in\hat{\CC}_{n,k}$, the delta distribution $\delta_C$ does not depend on $\te_{k+1}$. The function $S_C^{(k)}$ depends on $\te_{k+1}$ only via $m_j=k+1$ in $S_{m_j,r_j}^{(k)}$ in \eqref{def-dS} because $l_i,r_j\neq k+1$. Since $S_2$ is analytic in $S(0,\pi)$, the factor $S_{k+1,r_j}^{(k)} = S_{r_j,k+1}$ has an analytic continuation in $\te_{k+1}$ to the strip $S(-\pi,0)$, with the crossing-symmetric boundary value $S_2(\te_{r_j}-(\te_{k+1}-i\pi))=S_2(\te_{k+1}-\te_{r_j})=S_2^{(k+1)}(\te_{k+1}-\te_{r_j})$. All other factors in $S_C^{(k)}$ are of the form $S_{a,b}^{(k)}$, $a,b\neq k+1$, and therefore satisfy $S_{a,b}^{(k)}=S_{a,b}^{(k+1)}$. Thus $\te_{k+1}\mapsto S_C^{(k)}(\bte)$, with $\te_1,...,\te_k,\te_{k+2},...,\te_n\in\Rl$ fixed, can be analytically continued to $S(-\pi,0)$, with boundary value $S_C^{(k+1)}$ at $\Rl-i\pi$.

According to Lemma \ref{lemma:chat}, also $\langle \bl_C\cup\{k+1\}|\,[z_{k+1},A]\,|\br_C\rangle_{n,k}$ has an analytic continuation in $\te_{k+1}\in S(-\pi,0)$, and its boundary value at Im$\,\te_{k+1}=-\pi$ is obtained by exchanging $[z_{k+1},A]$ with $[A,\zd_{k+1}]$. Hence $\langle A\rangle^{\rm con}_{n,k}$ has an analytic continuation to the strip $S(-\pi,0)$, and its boundary value at Im$\,\te_{k+1}=-\pi$ is (in the sense of distributions)
\begin{align*}
\langle A\rangle^{\rm con}_{n,k}(\te_1,..,\te_{k+1}-i\pi,..,\te_n)
&=
\sum_{C\in\hat{\CC}_{n,k}}
\delta_C S_C^{(k+1)}
\langle \bl_C\cup\{k+1\}|\,[A,\zd_{k+1}]\,|\br_C\rangle_{n,k}\,.
\end{align*}
Taking into account the formula \eqref{acon-comm2} for $\langle A\rangle^{\rm con}_{n,k}$, this shows that the boundary value of $\langle A\rangle^{\rm con}_{n,k}$ at Im$\,\te_{k+1}=-\pi$ is $\langle A\rangle^{\rm con}_{n,k+1}$.

b) Let $C\in\hat{\CC}_{n,k}$ and put $\bte_\sbr := (\te_{r_1},...,\te_{r_{|C|}})$. In the product $S_C^{(k)}$ \eqref{def-dS}, at least one of the two variables $\te_a,\te_b$ of each factor $S_{a,b}$ is contracted, i.e. either $a\in\bl_C\cup\br_C$ or $b\in\bl_C\cup\br_C$. After the multiplication with $\delta_C$, the variables $\te_{l_j}$ and $\te_{r_j}$ are identified. We can therefore split $\delta_C S_C^{(k)}$ into a product of three factors, $\delta_C S_C^{(k)}= \delta_C S_C^LS_C^MS_C^R$, where $S_C^L$ depends on $\{\te_{k+2},...,\te_n\}\backslash\{\te_{l_1},...,\te_{l_{|C|}}\}$ and $\bte_\sbr$, and $S_C^R$ depends on $\{\te_1,...,\te_k\}$. Only $S_C^M=\prod_{j=1}^{|C|}S_{r_j,k+1}$ depends on $\te_{k+1}$.

For $f_1,...,f_n\in\Ss(\Rl)$, let
\begin{align}
F^L_{\sbte_\sbr}
&:=
S_C^L \cdot 
\big(
	f_{k+2}\otimes ... \otimes \widehat{f_{l_1}}\otimes ... \otimes \widehat{f_{l_{|C|}}}\otimes ... \otimes f_n
\big)\,,\label{def:FL}
\\
F^R_{\sbte_\sbr}
&:=
S_C^R \cdot 
\big(
	f_k\otimes ... \otimes \widehat{f_{r_1}}\otimes ... \otimes \widehat{f_{r_{|C|}}}\otimes ... \otimes f_1
\big)
\,,
\end{align}
where the hats indicate omission of the corresponding factors. $F^L_{\sbte_\sbr}$ and $F^R_{\sbte_\sbr}$ are considered as functions of the $n-k-1-|C|$ variables $\{\te_{k+2},...,\te_n\}\backslash\{\te_{l_1},...,\te_{l_{|C|}}\}$ and the $k-|C|$ variables $\{\te_1,...,\te_k\}\backslash\{\te_{r_1},...,\te_{r_{|C|}}\}$, respectively, and the dependence of these functions on $\bte_\sbr\in\Rl^{|C|}$ is treated as a parameter. In view of $|S_2(\te)|=1$, $\te\in\Rl$, the $L^2$-norms of $F^{L/R}_{\sbte_\sbr}$ are
\begin{align}\label{bnd:FLR}
\|F^L_{\sbte_\sbr}\| 
=
\prod_{\genfrac{}{}{0pt}{}{j=k+2}{j\notin \sbl_C}}^n \|f_j\|_2
\,,\qquad
\|F^R_{\sbte_\sbr}\| 
=
\prod_{\genfrac{}{}{0pt}{}{j=1}{j\notin \sbr_C}}^k \|f_j\|_2
\,,\qquad
\bte_\sbr\in\Rl^{|C|}\,.
\end{align}
Let
\begin{align*}
\langle A \rangle^{\rm con}_{n,k}(\bof;\te_{k+1})
&:=
 \int \langle A \rangle^{\rm con}_{n,k}(\te_1,...,\te_{k+1},...,\te_n)\prod_{\genfrac{}{}{0pt}{}{j=1}{j\neq k+1}}^n f_j(\te_j)\,d\te_j \,.
\end{align*}
After analytic continuation in $\te_{k+1}$, and after carrying out the integration over the delta distributions in \eqref{acon-comm1}, we find
\begin{align*}
\langle A \rangle^{\rm con}_{n,k}(\bof;\te_{k+1}-i\la)
&=
\sum_{C\in\hat{\CC}_{n,k}}(-1)^{|C|}
\int d^{|C|}\bte_\sbr \,
I^C_{\sbte_\sbr,\la}(\te_{k+1})\prod_{j=1}^{|C|} f_{l_j}(\te_{r_j})f_{r_j}(\te_{r_j})
\,\\
I^C_{\sbte_\sbr,\la}(\te_{k+1})
= \prod_{j=1}^{|C|} &
S_2(\te_{r_j}-\te_{k+1}+i\la)
\;\times\\
&\times
\langle\bl_C\cup\{k+1\}|\,[z(\te_{k+1}-i\la),A]\,|\br_C\rangle_{n,k}
(F^L_{\sbte_\sbr} \otimes F^R_{\sbte_\sbr})
\,.
\end{align*}
Putting the bounds $|S_2(\zeta)|\leq 1$, $\zeta\in S(0,\pi)$, \eqref{Chardybound}, \eqref{lr-bnd} and \eqref{bnd:FLR} together, we arrive at, $0\leq\la\leq\pi$,
\begin{align}
&\left|\int_\Rl d\te_{k+1}f_{k+1}(\te_{k+1}) I^C_{\sbte_\sbr,\la}(\te_{k+1})\right|
 \leq \gamma^C_{n,k}\cdot \prod_{j\neq\sbl_C\cup\sbr_C}\|f_j\|_2\cdot\|A\|\,,
 \label{bnd-IC}
 \\
 \gamma^C_{n,k} &:=
 \left(\sqrt{n-k-|C|}+\sqrt{k-|C|+1}\right)\sqrt{(n-k-1-|C|)!(k-|C|)!}\nonumber
  \leq \frac{2\sqrt{n!}}{|C|!}
 .
\end{align}
The given estimate on $\gamma^C_{n,k}$ follows from the inequality $a!b!\leq (a+b)!$.
From \eqref{bnd-IC} we conclude 
\begin{align}\label{bnd-fast}
 \left|\int d\te_{k+1}\,f_{k+1}(\te_{k+1})\langle A \rangle^{\rm con}_{n,k}(\bof;\te_{k+1}-i\la)\right|
 \leq
 2\sqrt{n!}\sum_{C\in\hat{\CC}_{n,k}}\frac{\|A\|}{|C|!}\prod_{j=1}^n\|f_j\|_2.
\end{align}
It remains the combinatorial problem to find a bound on the sum over $\hat{\CC}_{n,k}$. Note that the number of all contractions $C\in\hat{\CC}_{n,k}$ with fixed length $|C|=N$ is $N!\genfrac{(}{)}{0pt}{}{k}{N}\genfrac{(}{)}{0pt}{}{n-k-1}{N}$, since each such contraction is given by two $N$-element subsets $\{r_1,...,r_N\}\subset\{1,...,k\}$ and $\{l_1,...,l_N\}\subset\{k+2,...,n\}$, and a permutation of $\{1,...,N\}$ to determine which element of $\{l_1,...,l_N\}$ is contracted with which element of $\{r_1,...,r_N\}$. Using $|C|\leq\min\{k,n-k-1\}$, we find
\begin{align*}
\sum_{C\in\hat{\CC}_{n,k}}\frac{1}{|C|!}
&=
\sum_{N=0}^{\min\{k,n-k-1\}}\genfrac{(}{)}{0pt}{}{k}{N}\genfrac{(}{)}{0pt}{}{n-k-1}{N}
\\
&\leq
\sum_{N=0}^k \sum_{M=0}^{n-k-1} \genfrac{(}{)}{0pt}{}{k}{N}\genfrac{(}{)}{0pt}{}{n-k-1}{M}
=
2^{n-1}\,.
\end{align*}
In combination with \eqref{bnd-fast}, this implies the desired bound \eqref{acon-bnd}.
\end{proof}
The analyticity and boundedness properties of the contracted matrix elements $\langle A\rangle_{n,k}^{\rm con}$ imply corresponding properties of the $n$-particle form factors $(A\Om)_n$. In order not to overburden our notation, we will denote the analytic continuation of $(A\Om)_n$ by the same symbol. In the following, more specific information on the underlying regular scattering function is needed. We will exploit the fact that each $S_2\in\SF_0$ can be continued to the enlarged strip $S(-\kappa(S_2),\pi+\kappa(S_2))$, and is bounded by $\|S_2\|<\infty$ on this domain. Also recall that $\kappa(S_2)\leq\frac{\pi}{2}$ by definition.

\begin{figure}[here]
    \leavevmode
    \begin{picture}(0,0)%
\centering\epsfig{file=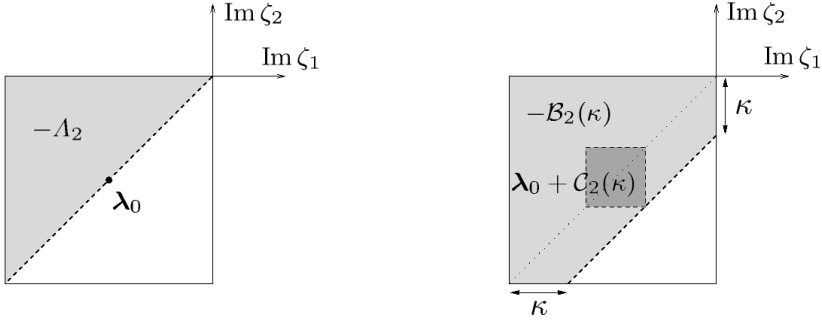,width=12cm}
\end{picture}%
 \setlength{\unitlength}{2763sp}%
 \begin{picture}(5476,2807)(3263,-3061)
 \end{picture}
    \caption{The two-dimensional bases $-\La_2$ (triangle on the left), $-\Ba_2(\kappa)$ (pentagon on the right) and $\bla_0+\Cu_2(\kappa)$ (square on the right), inscribed in the square $[-\pi,0]\times[-\pi,0]$.}
    \label{fig:bases}
\end{figure}

The regions which are relevant in this context are, $\kappa>0$,
 \begin{align}\label{def:Tn}
      \Lambda_n
      &:=
      \left\{\bla\in\Rl^n \,:\, \pi>\la_1>\la_2>...>\la_n>0\right\}\,.
\\    \label{bibobase}
      \Ba_n(\kappa)
      &:= 
      \big\{\bla\in \Rl^n
      \,:\, 
      0<\la_1,...,\la_n < \pi,\;\,
      \la_k-\la_l < \kappa,\,
      1\leq l< k\leq n
      \big\}\,,\\
\bla_0&:=\left(-\frac{\pi}{2},...,-\frac{\pi}{2}\right)
\,,\qquad
\Cu_n(\kappa):=\left(-\frac{\kappa}{2},\frac{\kappa}{2}\right)^{\times n}\,.
\end{align}
Note that $\Cu_n(\kappa)+\bla_0 \subset \Ba_n(\kappa)$ (cf. figure \ref{fig:bases} for the case $n=2$). The tubes based on these sets are denoted
  \begin{align}
      \Tu_n&:=\Rl^n-i\Lambda_n \,,\\
\Tu_n(\kappa)
&:=
\Rl^n + i\big(\bla_0+\Cu_n(\kappa)\big)
\,.\label{def:tkn}
\end{align}
\begin{proposition}{\bf(Analyticity and boundedness properties of  \mbox{\boldmath{$(A\Om)_n$}})}\label{prop:ana}\\
 In a model with regular scattering function, let $A\in\M$.
  \begin{enumerate}
  \item $(A\Omega)_n$ is analytic in the tube $\Rl^n-i\Ba_n(\kappa(S_2))$.
\item Let $0<\kappa<\kappa(S_2)$. There holds the bound, 
    \begin{align}\label{masterbound}
      |(A\Omega)_n(\bze)|
      &\leq
	\left(\frac{8}{\pi}\frac{\|S_2\|}{\sqrt{\kappa(S_2)-\kappa}}\right)^n\cdot\|A\|
      \,,\qquad\bze\in\Tu_n(\kappa)\,.
    \end{align}
  \end{enumerate}
\end{proposition}
\begin{proof}
a) Let $\bof\in\Ss(\Rl^n)$. As the first statement to be proven, we claim that the convolution $(\te_1,...,\te_k)\mapsto((A\Omega)_n*\bof)(\te_1,...,\te_n)$, considered as a function of $\te_1,...,\te_k$, with $\te_{k+1},...,\te_n\in\Rl$ fixed, is analytic in the tube $\Rl^k-i\Lambda_k$ and continuous on its closure. Our proof is based on induction in $k\in\{1,...,n\}$.

For $k=1$, note that $(A\Om)_n=\langle A\rangle^{\rm con}_{n,0}/\sqrt{n!}$, since $\CC_{n,0}$ contains only the empty contraction \eqref{def-Acon}. But according to Lemma \ref{lemma:recursion} a), $\langle A\rangle^{\rm con}_{n,0}$ is the boundary value of a function analytic in $S(-\pi,0)=\Rl^1-i\Lambda_1$. Thus the claim for $k=1$ follows.

So assume analyticity of $(\te_1,...,\te_k)\longmapsto ((A\Omega)_n*\bof)(\te_1,...,\te_n)$ in $\Rl^k-i\Lambda_k$. In view of Lemma \ref{lemma:recursion} a), the boundary value at ${\rm Im}\,\te_1=...={\rm Im}\,\te_k=-\pi$ is given by $\langle A\rangle^{\rm con}_{n,k}*\bof/\sqrt{n!}$, which in turn has an analytic continuation in $\te_{k+1}\in S(-\pi,0)$. By application of the Malgrange Zerner (``flat tube'') theorem (cf., for example, \cite{Epstein66}), it follows that $(A\Omega)_n*\bof$, considered as a function of the first $k+1$ variables, has an analytic continuation to the convex closure of the set
\begin{align*}
  \Rl^{k+1}
  -i \big(
  \left\{(\la_1,..,\la_k,0)\,:\,(\la_1,..,\la_k)\in\Lambda_k\right\}
  \cup
  \left\{(\pi,..,\pi,\la_{k+1})\,:\, \pi > \la_{k+1} > 0 \right\}\!\big)
  \,,
\end{align*}
which coincides with $\Rl^{k+1}-i\Lambda_{k+1}$. Hence our claim follows. Since $\bof$ was arbitrary, we conclude  that $(A\Omega)_n$ is the boundary value in the sense of distributions of a function (denoted by the same symbol) analytic in $\Tu_n=\Rl^n-i\La_n$ \eqref{def:Tn}.
\\\\
Now let $\frS_n$ denote the group of permutations of $n$ objects and consider the ``permuted form factors'' 
\begin{align*}
  (A\Omega)_n^\rho(\bte) 
  &:=
  (A\Omega)_n(\te_{\rho(1)},...,\te_{\rho(n)})
  \,,\qquad \rho\in\frS_n\,,
\end{align*}
 which are analytic in the ``permuted tubes'' $\Tu_n^\rho :=\Rl^n-i\Lambda_n^\rho$\,,
\begin{align*}
  \Lambda_n^\rho
  &:=
  \big\{\bla\in\Rl^n\,:\,  \pi>\la_{\rho(1)}>...>\la_{\rho(n)}>0\big\}
  \;.
\end{align*}
Recall that $(A\Om)_n\in\Hil_n$ is invariant under the representation $D_n$ of $\frS_n$ \eqref{def:Dn}, 
\begin{align}\label{psi-perm}
  (A\Omega)_n(\bte)
  =
  (D_n(\rho)(A\Om)_n)(\bte)
  &=
  S^\rho(\bte)\cdot(A\Omega)_n^\rho(\bte)
  \,,\\
  S^\rho(\bte)
  &=
  \prod_{\genfrac{}{}{0pt}{}{1\leq l < k \leq n}{\rho(l) > \rho(k)}}
  S_2(\te_{\rho(l)}-\te_{\rho(k)})
  \;.\label{srho}
\end{align} 
As $S_2\in\SF_0$ is analytic in $S(-\kappa(S_2),\pi+\kappa(S_2))$, all the functions $S^\rho$, $\rho\in\frS_n$, are
analytic in the tube $\Rl^n + i \,\Ba'_n(\kappa(S_2))$ with base
\begin{align*}
  \Ba'_n(\kappa(S_2))
  &:=
  \big\{\bla\in\Rl^n
  \,:\, 
  -\kappa(S_2) < \la_k-\la_l < \pi + \kappa(S_2),
  \quad 1\leq l<k\leq n
  \big\}\,.
\end{align*}
Hence the right hand side of \eqref{psi-perm} can be analytically continued to the tube based on $\Ba'_n(\kappa(S_2))\cap(-\Lambda_n^\rho)$. But the left hand side of \eqref{psi-perm} is analytic in $\Rl^n-i\Lambda_n$, and both sides converge in the sense of distributions to the same boundary values on $\Rl^n$. So we may apply Epstein's generalization of the Edge of the Wedge Theorem \cite{epstein-eotw} to conclude that $(A\Omega)_n$ has an analytic continuation to the tube whose base is the convex closure of 
\begin{align*}
  \bigcup_{\rho\in\frS_n} \Ba'_n(\kappa(S_2))\cap(-\Lambda_n^\rho)\,.
\end{align*}
Since the convex closure of $\bigcup_{\rho}\Lambda_n^\rho$ is the cube $(0,\pi)^{\times n}$, it follows that $(A\Omega)_n$ is
analytic in the tube based on $(-\pi,0)^{\times n}\cap\Ba'_n(\kappa(S_2))=-\Ba_n(\kappa(S_2))$, and the proof of part a) is finished.
\\
\\
b) We first derive an estimate on $|S^\rho(\bze)|$ \eqref{srho}, $\bze\in\Tu_n(\kappa(S_2))$. Clearly, $S^\rho$ is bounded on $\Tu_n(\kappa(S_2))$, because each factor $S_2(\zeta_{\rho(l)}-\zeta_{\rho(k)})$ is bounded. By the multidimensional analogue of the three lines theorem \cite{bochner-martin}, the supremum of $S^\rho$ over this tube is attained on a subspace of the form $\Rl^n+i\bla_0+i\bxi$, where $\bxi$ is a vertex of $\Cu_n(\kappa(S_2))$, i.e.
\begin{align*}
 |S^\rho(\bze)| \leq \sup_{\sbte\in\Rl^n}\prod_{\genfrac{}{}{0pt}{}{1\leq l < k \leq n}{\rho(l) > \rho(k)}}
  |S_2(\te_{\rho(l)}-\te_{\rho(k)}+i(\xi_{\rho(l)}-\xi_{\rho(k)}))|
\,,\qquad\bze\in\Tu_n(\kappa(S_2))
\,.
\end{align*}
Since $\bxi$ is a vertex of $\Cu_n(\kappa(S_2))$, there holds $\xi_{\rho(l)}-\xi_{\rho(k)}\in\{0,\kappa(S_2),-\kappa(S_2)\}$. We have
\begin{align*}
\sup_{\te\in\Rl}|S_2(\te)|=1\,,\qquad \sup_{\te\in\Rl}|S_2(\te+i\kappa(S_2))|\leq 1\,,\qquad \sup_{\te\in\Rl}|S_2(\te-i\kappa(S_2))|\leq\|S_2\|\,.
\end{align*}
As at most $(n-1)$ of the differences $\xi_{\rho(l)}-\xi_{\rho(k)}$ can equal $-\kappa(S_2)$ simultaneously, and since $\|S_2\|\geq 1$, we conclude 
\begin{align}\label{bnd-Srho}
|S^\rho(\bze)| \leq \|S_2\|^{\,n-1} \leq \|S_2\|^n\,,\qquad\bze\in\Tu_n(\kappa(S_2))\,.
\end{align}

Now let $f_1,...,f_n\in \Ss(\Rl)$ and put $\bof:= f_1\otimes ... \otimes f_n$. Lemma \ref{lemma:recursion} b) implies that at points $\bte-i\bla\in\overline{\Tu_n}$ with $\bla=(\pi,...,\pi,\la_{k+1},0,...,0)$,  $0\leq \la_{k+1}\leq\pi$, there holds the bound
\begin{align}\label{bnd34}
  |((A\Omega)_n*\bof)(\bte-i\bla)|
  \leq
  2^n\|A\|\,\prod_{j=1}^n\|f_j\|_2\,.
\end{align}
By a standard argument (cf., for example \cite[Lemma A.2]{GL-bros}), this bound can be seen to hold for arbitrary $\bla\in\overline{\Lambda_n}$. Moreover, taking into account the $S_2$-symmetry of $(A\Om)_n$ \eqref{psi-perm} and the bound \eqref{bnd34}, we find
\begin{align}\label{bnd34-2}
  |((A\Omega)_n*\bof)(\bze)|
  \leq
  \left(2\|S_2\|\right)^n\|A\|\,\prod_{j=1}^n\|f_j\|_2\,,\qquad \bze\in\Tu_n(\kappa(S_2))\,,
\end{align}
and this inequality extends to $f_1,...,f_n\in L^2(\Rl)$ by continuity. 

To proceed to the desired bound \eqref{masterbound}, we fix arbitrary $\bte\in\Rl^n$, $\bla\in\bla_0+\Cu_n(\kappa)$, and $0<\kappa<\kappa(S_2)$, put $\bte-i\bla=:\bze\in\Tu_n(\kappa)$, and consider the disc $\Di_r(\zeta_k)\subset\Cl$ of radius $r:=\frac{1}{2}(\kappa(S_2)-\kappa)$ and center $\zeta_k$. For this value of the radius, the polydisc $\Di_r(\zeta_1)\times...\times\Di_r(\zeta_n)$ is contained in $\Tu_n(\kappa(S_2))$. Denoting the characteristic function of the interval $[-r(\la_k'),r(\la_k')]$, with $r(\la_k'):=\sqrt{r^2-(\la_k')^2}$, by $\chi_{\la_k'}$, we can use the mean value property for analytic functions as follows.
\begin{align*}
(A\Om)_n(\bze)
&=
(\pi r^2)^{-n} \int\limits_{\Di_r(\zeta_1)}\!\!d\te'_1 d\la'_1\cdots \int\limits_{\Di_r(\zeta_n)}\!\! d\te_n' d\la_n' \; (A\Om)_n(\bte'+i\bla')
\\
 &=
 (\pi r^2)^{-n} \!\!\! \int\limits_{[-r,r]^{\times n}} \!\!\!\!d^n\bla'\!\int\limits_{-r(\la_1')}^{r(\la_1')}\!\!d\te_1' \cdots\!\!\!\! \int\limits_{-r(\la_n')}^{r(\la_n')}\!\!d\te_n'
 \,(A\Om)_n(\bte+\bte'-i\bla+i\bla')
 \\
&=
(\pi r^2)^{-n} \int\limits_{[-r,r]^{\times n}} d^n\bla'\, \left( (A\Om)_n * (\chi_{\la_1'}\otimes ... \otimes \chi_{\la_n'})\right) (\bte-i\bla+i\bla')
\,.
\end{align*}
Since $\bte-i\bla+i\bla'\in\Tu_n(\kappa(S_2))$, we can apply the estimate \eqref{bnd34-2}. Taking into account $\|\chi_{\la_k'}\|_2=\sqrt{2r(\la_k')}\leq\sqrt{2r}$ and $r=\frac{1}{2}(\kappa(S_2)-\kappa)$, we get
\begin{align*}
\left|(A\Om)_n(\bze)\right|
&\leq
(\pi r^2)^{-n}\cdot(2r)^n(2\|S_2\|)^n\|A\|\cdot (2r)^{n/2}
=
	\left(
		\frac{8}{\pi}\frac{\|S_2\|}{\sqrt{\kappa(S_2)-\kappa}}
	\right)^n
	\|A\|.
\end{align*}
Since $\bze\in\Tu_n(\kappa)$ was arbitrary, the proof is finished.
\end{proof}

\section{Proof of the Nuclearity Condition}\label{sec:nuc}

With the help of the results of the previous section, we can now proceed to the proof of the modular nuclearity condition. We begin by recalling the definition of a nuclear map \cite{jarchow,pietsch}.
\begin{definition}{\bf(Nuclear maps)}\label{def:nucmap}\\
Let $\X$ and $\Y$ be two Banach spaces. A linear map $T:\X\longrightarrow\Y$ is said to be nuclear if there exists a sequence of vectors $\{\Psi_k\}_k\subset\Y$ and a sequence of linear functionals $\{\eta_k\}_k\subset\X_*$ such that, $X\in\X$,
	\begin{align}\label{nuc-decc}
		T(X) &= \sum_{k=1}^\infty \eta_k(X)\,\Psi_k\,,\qquad
		\sum_{k=1}^\infty \|\eta_k\|_{\X_*} \|\Psi_k\|_\Y < \infty \,.
	\end{align}
	The nuclear norm of such a mapping is defined as
	\begin{align}\label{def:nuclearnorm}
		\|T\|_1 &:= \inf_{\eta_k,\Psi_k} \sum_{k=1}^\infty \|\eta_k\|_{\X_*} \|\Psi_k\|_\Y\,,
	\end{align}
	where the infimum is taken over all sequences $\{\Psi_k\}_k\subset\Y$, $\{\eta_k\}_k\subset\X_*$  complying with the above conditions.
\end{definition}
The sets of all bounded respectively nuclear maps between $\X$ and $\Y$ will be denoted $\B(\X,\Y)$ and $\NN(\X,\Y)$, respectively.
We will use the following well-known facts about nuclear maps, mostly without further mentioning.
\\
\begin{lemma}{\bf(Properties of nuclear maps)}\label{lemma:nucmaps}\\
 Let $\X,\X_1,\Y,\Y_1$ be Banach spaces.
 \begin{enumerate}
 \item Let $A_1\in\B(\X,\X_1)$, $T\in\NN(\X_1,\Y_1)$, $A_2\in\B(\Y_1,\Y)$. Then $A_2TA_1\in\NN(\X,\Y)$, and
 \begin{equation}
 \|A_2TA_1\|_1\leq\|A_2\|\cdot\|T\|_1\cdot\|A_1\|\,.
\end{equation}
\item $(\NN(\X,\Y),\|\cdot\|_1)$ is a Banach space.
\item Let $\Hil$ be a separable Hilbert space. Then $\NN(\Hil,\Hil)$ coincides with the set of trace class operators on $\Hil$, and 
\begin{equation}
 \|T\|_1={\rm Tr}\,|T|\,,\qquad T\in\NN(\Hil,\Hil)\,.
\end{equation}
\end{enumerate}
\end{lemma}
For a proof of this lemma, see for example \cite{jarchow}.

We also have to recall the notion of Hardy spaces on tubes.
\begin{definition}{\bf(Hardy spaces on tube domains)}\\
Let $\Cu\subset\Rl^n$ be open. The Hardy space $H^2(\Tu)$ on the tube $\Tu=\Rl^n+i\,\Cu$ is the space of all analytic functions $F:\Tu\to\Cl$ for which $F_\sbla$ is an element of $L^2(\Rl^n)$ for each $\bla\in\Cu$, and which have finite Hardy norm
\begin{align}
\bno{F}
&:=
\sup_{\sbla\in\Cu}\|F_\sbla\|_2
=
\sup_{\sbla\in\Cu}\left(\int_{\Rl^n} d^n\bte \,|F(\bte+i\bla)|^2\right)^{1/2}
<
\infty\,.
\end{align}
\end{definition}
$(H^2(\Tu),\bno{\cdot})$ is a Banach space \cite{StW}. 

As in the preceding section, we choose $\kappa$ in $0<\kappa<\kappa(S_2)\leq\frac{\pi}{2}$, and consider the tube
\begin{align*}
 \Tu_n(\kappa) := \bla_0 +i\,\Cu_n(\kappa)\,,\qquad 
\bla_0 := -\left(\frac{\pi}{2},...,\frac{\pi}{2}\right)\,,\qquad \Cu_n(\kappa):=\left(-\frac{\kappa}{2},\frac{\kappa}{2}\right)^{\times n}\,. 
\end{align*}

Having set up our notation, we now turn to the analysis of the properties of the concrete mappings $\Xi_n(s)$ \eqref{def:xins} appearing in the modular nuclearity condition, 
\begin{align}
\Xi_n(s):\M&\longrightarrow\Hil_n\subset L^2(\Rl^n)\,,\qquad s>0\,,\nonumber
\\
 \left(\Xi_n(s)A\right)_n(\te_1,...,\te_n)
 &=
 \prod_{k=1}^n e^{-ms\cosh\te_k}\cdot (A\Om)_n(\te_1-\tfrac{i\pi}{2},...,\te_n-\tfrac{i\pi}{2})\,,\label{def:xins-2}
\end{align}
We decompose $\Xi_n(s)$ as depicted in the following diagram:
\begin{diagram}
\M &   &  \\
\dTo^{\Sigma_n(s,\kappa)} & \rdTo^{\Xi_n(s)} &  \\
H^2(\Tu_n(\kappa)) & \rTo_{\Delta_n(s,\kappa)} & L^2(\Rl^n) \\
\end{diagram}
 The two maps $\Sigma_n(s,\kappa):\M \to H^2(\Tu_n(\kappa))$ and $\Delta_n(s,\kappa):H^2(\Tu_n(\kappa))\to L^2(\Rl^n)$ appearing here are defined as
\begin{align}
 \Sigma_n(s,\kappa)A 		&:=(A(\tfrac{1}{2}\us)\Om)_n\,,\qquad \us:=(0,s)\,,\label{def:Sigma_n}\\
(\Delta_n(s,\kappa)F)(\bte)	&:=\prod_{k=1}^n e^{-\frac{ms}{2}\cosh\te_k}\cdot F(\bte+i\bla_0)\,.\label{def:Delta_n}
\end{align}
In view of \eqref{def:xins-2}, the above diagram commutes, i.e. there holds
\begin{align}
 \Xi_n(s)A	&= \Delta_n(s,\kappa)\Sigma_n(s,\kappa)A\,,\qquad A\in\M\,.
\end{align}
$\Sigma_n(s,\kappa)$ and $\Delta_n(s,\kappa)$ are investigated in the following two Lemmas.
\begin{lemma}\label{cor:hardy}
Let $S_2\in\SF_0$ and $0<\kappa<\kappa(S_2)$. The map $\Sigma_n(s,\kappa)$, $s>0$, is a bounded operator between the Banach spaces $(\M,\|\cdot\|_{\B(\Hil)})$ and $(H^2(\Tu_n(\kappa)),\bno{\cdot})$. Its operator norm satisfies
\begin{align}\label{sigma-bound}
\|\Sigma_n(s,\kappa)\| \leq
\sigma(s,\kappa)^n\,.
\end{align}
For fixed $\kappa$, the function $s\mapsto\sigma(s,\kappa)$ is monotonously decreasing, with the limits $\sigma(s,\kappa)	\to0$ for $s\to\infty$ and $\sigma(s,\kappa)\to\infty$ for $s\to 0$.
\end{lemma}
\begin{proof}
Given the translation invariance of $\Om$ and the form of $U$ \eqref{def:U}, we have
\begin{align*}
 (\Sigma_n(s,\kappa)A)(\bze) &= (A(\tfrac{1}{2}\us)\Om)_n(\bze)	= u_{n,s}(\bze)\cdot(A\Om)_n(\bze)\,,\\
u_{n,s}(\bze)&=\prod_{k=1}^n e^{-\frac{ims}{2}\sinh\zeta_k}\,.
\end{align*}
Since $u_{n,s}$ is entire, the analyticity of $(A\Om)_n$ (Proposition \ref{prop:ana}) carries over to $\Sigma_n(s,\kappa)A$. Moreover, it follows from a straightforward calculation that $u_{n,s}$ is an element of $H^2(\Tu_n(\kappa))$, with Hardy norm
\begin{align}\label{un-hardy-bound}
\bno{u_{n,s}}
=
\left(\int_\Rl d\te\, e^{-ms\cos\kappa\,\cosh\te}\right)^{n/2}\,,
\end{align}
and this integral converges since $s>0$ and $0<\kappa<\frac{\pi}{2}$.

In view of the uniform bound \eqref{masterbound} on $(A\Om)_n(\bze)$, $\bze\in\Tu_n(\kappa)$, it follows that also $u_{n,s}\cdot(A\Om)_n$ lies in the Hardy space $H^2(\Tu_n(\kappa))$, with norm bounded by
\begin{align}
\frac{\bno{\Sigma_n(s,\kappa)A}}{\|A\|} &\leq \left(\frac{8}{\pi}\frac{\|S_2\|}{\sqrt{\kappa(S_2)-\kappa}}
\cdot
\left(\int_\Rl d\te\, e^{-ms\cos\kappa\,\cosh\te}\right)^{1/2}\right)^n\,.
\end{align}
The claimed behaviour of $\|\Sigma_n(s,\kappa)\|$ with respect to $s$ can be directly read off from this formula.
\end{proof}

\begin{lemma}\label{lemma:delta}
  Let $s>0$, $\kappa > 0$, and $\Delta_n(s,\kappa)$ be defined as in \eqref{def:Delta_n}.
  \begin{enumerate}
  \item $\Delta_n(s,\kappa)$ is a nuclear
    map between the Banach spaces $(H^2(\Tu_n(\kappa)),\bno{\cdot})$
    and $(L^2(\Rl^n),\|\cdot\|_2)$.
  \item Let $T_{s,\kappa}$ be the integral operator on $L^2(\Rl,d\te)$ with kernel
  \begin{eqnarray}\label{int-kernel2}
    T_{s,\kappa}(\te,\te')
    &=&
    \frac{e^{-\frac{ms}{2}\cosh\te}}
    {i\pi \,(\te'-\te - \frac{i\kappa}{2})}   
     \;.
   \end{eqnarray}
   $T_{s,\kappa}$ is of trace class, and there holds the bound
  \begin{eqnarray}\label{delta-bound}
    \|\Delta_n(s,\kappa)\|_1
    &\leq&
    \|T_{s,\kappa}\|_1^{\,n}
    < \infty
    \,.
  \end{eqnarray}
  \end{enumerate}
 \end{lemma}
\begin{proof}
Let $F\in H^2(\Tu_n(\kappa))$, and pick $\bte\in\Rl^n$ and a polydisc $\Di_n(\bte+i\bla_0)\subset\Tu_n(\kappa)$ with center $\bte+i\bla_0$. By virtue of Cauchy's integral formula, we can represent $F(\bte+i\bla_0)$ as a contour integral over $\Di_n(\bte+i\bla_0)$,
\begin{align}\label{contourint}
    F(\bte+i\bla_0)
    &=
    \frac{1}{(2\pi i)^n}
    \oint\limits_{\Di_n(\sbte+i\sbla_0)} d^n\bze'\;
    \frac{F(\bze')}{\prod_{k=1}^n(\zeta'_k-\te_k + \frac{i\pi}{2})}
    \;\;.
\end{align}
As a consequence of the mean value property, the Hardy space function $F$ is uniformly bounded on the subtubes $\Tu_n(\kappa')\subset\Tu_n(\kappa)$, $\kappa'<\kappa$ (cf. the line of argument at the end of the proof of Proposition \ref{prop:ana} b)).

Moreover, $F\in H^2(\Tu_n(\kappa))$ can be continued to the boundary of $\Tu_n(\kappa)$ as follows: The map $\Cu_n(\kappa)\ni\bla\mapsto F_\sbla\in L^2(\Rl^n)$ extends continuously (in the norm topology of $L^2(\Rl^n)$) to the closed cube $\bla_0+[-\frac{\kappa}{2},\frac{\kappa}{2}]^{\times n}$ \cite[Ch. III, Cor. 2.9]{StW}.

Taking advantage of these two properties of $F$, we can deform the contour of integration in \eqref{contourint} to the boundary of $\Tu_n(\kappa)$. After multiplication with the exponential factor \eqref{def:Delta_n} we arrive at
\begin{align*}
    (\Delta_n(s,\kappa)F)(\bte)
    &=
    \frac{1}{(2\pi i)^n}\sum_\sbeps
    \int_{\Rl^n}d^n\bte'\,
    \prod_{k=1}^n \frac{\eps_k\;e^{-\frac{ms}{2}\cosh\te_k}}{\te_k'-\te_k-\frac{i\eps_k\kappa}{2}}
    \cdot F_{\sbla_0-\frac{\kappa}{2} \sbeps}(\bte')
    \;,
\end{align*}
where the summation runs over $\beps=(\eps_1,...,\eps_n)$, $\eps_1,...,\eps_n=\pm 1$. Expressed in terms of the integral operator $T_{s,\kappa}$, this equation reads
\begin{align}\label{delta-eqn}
    \Delta_n(s,\kappa)F
    &=
    2^{-n}\sum_{\sbeps} \eps_1\cdots\eps_n (T_{s,\eps_1\kappa}\otimes ... \otimes T_{s,\eps_n\kappa})F_{\sbla_0-\frac{\kappa}{2}\sbeps}\;.
\end{align}
The integral operators $T_{s,\pm\kappa}$ are of trace class on $L^2(\Rl)$, as can be shown by a standard argument \cite[Thm. XI.21]{SimonReed3}. Hence $T_{s,\eps_1\kappa}\otimes ... \otimes T_{s,\eps_n\kappa}$ is a trace class operator on $L^2(\Rl^n)$, for any $\eps_1,...,\eps_n=\pm 1$.  Note that since $T_{s,\kappa}$ and $T_{s,-\kappa}$ are related by $CT_{s,\kappa}C^*=-T_{s,-\kappa}$, $(Cf)(\te) := f(-\te)$, they have the same nuclear norm. Hence $\|T_{s,\eps_1\kappa}\otimes ... \otimes T_{s,\eps_n\kappa}\|_1=\|T_{s,\kappa}\|_1^n$.

Moreover, it follows from the $L^2$-convergence of $F$ to its boundary values that the maps $F\longmapsto F_{\sbla_0-\frac{\kappa}{2}\sbeps}$ are bounded as operators from $H^2(\Tu_n(\kappa))$ to $L^2(\Rl^n)$ for any $\beps$, with norm not exceeding one. According to Lemma \ref{lemma:nucmaps}, this implies the nuclearity of $\Delta_n(s,\kappa)$ \eqref{delta-eqn}. Since the sum in \eqref{delta-eqn} runs over $2^n$ terms, we also obtain the claimed bound $\|\Delta_n(s,\kappa)\|_1\leq \|T_{s,\kappa}\|_1^n$.
\end{proof}

Lemma \ref{lemma:delta} implies our first nuclearity result for the maps $\Xi(s)$ \eqref{def:XXi}.

\begin{theorem}{\bf (Nuclearity for sufficiently large splitting distances)}\label{thm:distalsplit}\\
In each model theory with regular scattering function, there exists a splitting distance $s_{\min}<\infty $ such that $\Xi(s)$ is nuclear for all $s>s_{\min}$.

Hence in these models, for each double cone $\OO_{a,b}=(W_R+a)\cap(W_L+b)$ with $b-a\in W_R$ and $-(b-a)^2>s_{\min}^2$, the corresponding observable algebra $\A(\OO_{a,b})=\M(a)\cap\M(b)'$ \eqref{def:AO} has $\Om$ as a cyclic vector.
\end{theorem}
\begin{proof}
  Let $\kappa\in(0,\kappa(S_2))$. 
  We have $\Xi_n(s) = \Delta_n(s,\kappa) \Sigma_n(s,\kappa)$, and in view of the previously established results, $\Xi_n(s)$ is nuclear, with nuclear norm bounded by (\ref{sigma-bound}, \ref{delta-bound}),
\begin{align}
    \|\Xi_n(s)\|_1 
    \;\leq\;
    \|\Sigma_n(s,\kappa)\|\cdot\|\Delta_n(s,\kappa)\|_1
    \;\leq\;
    \left(\sigma(s,\kappa)\cdot\|T_{s,\kappa}\|_1\right)^n
    \;.
\end{align}
To obtain nuclearity for $\Xi(s)=\sum_{n=0}^\infty \Xi_n(s)$ \eqref{def:Xin}, note that for $s\to\infty$, $\|T_{s,\kappa}\|_1$ and $\sigma(s,\kappa)$ converge monotonously to zero (cf. Lemma \ref{cor:hardy} and \eqref{int-kernel2}). So there exists $s_{\min}<\infty$ such that $\sigma(s,\kappa)\|T_{s,\kappa}\|_1 < 1$ for all $s>s_{\min}$. But for these values of $s$, there holds
\begin{align}
\sum_{n=0}^\infty \|\Xi_n(s)\|_1 
\leq
\sum_{n=0}^\infty \left(\sigma(s,\kappa)\,\|T_{s,\kappa}\|_1\right)^n
<
\infty
\,,
\end{align}
and the series $\sum_{n=0}^\infty \Xi_n(s)$ converges in nuclear norm to $\Xi(s)$. Since the set of nuclear operators between two Banach spaces is closed with respect to convergence in $\|\cdot\|_1$, the nuclearity of $\Xi(s)$ follows.

The Reeh-Schlieder property for the double cone algebras $\A(\OO_{0,\us})$ is a consequence of the nuclearity of $\Xi(s)$ (Theorem \ref{thm2}). 

For a region of the form $\OO_{0,x}$, $x\in W_R$, $-x^2>s_{\min}^2$, there exists a rapidity parameter $\la$ such that
\begin{equation}
 \left(
 \begin{array}{ll}
 \cosh\la&\sinh\la\\
 \sinh\la&\cosh\la
\end{array}
\right)
x=
\left(
\begin{array}{l}
 0\\s
\end{array}
\right)\,,\qquad s>s_{\min}\,.
\end{equation}
Since the modular operator of $(\M,\Om)$ commutes with the boosts $U(0,\la)$, and $U(0,\la)\Om=\Om$, it follows that $\Xi(x)$ and $\Xi(s)$ are related by
\begin{equation}
 \Xi(x)=U(0,\la)\Xi(s)\alpha_\la^{-1}\,,\qquad \alpha_\la(A):=U(0,\la)A\,U(0,\la)^{-1}\,.
\end{equation}
So the invariance of $\M$ under the modular group $\alpha_\la$ and the unitarity of $U(0,\la)$ imply that $\Xi(x)$ is nuclear, too, with $\|\Xi(x)\|_1=\|\Xi(s)\|_1$.

The corresponding statement for double cone regions $\OO_{a,b}$ with $b-a\in W_R$, $-(b-a)^2>s_{\min}^2$, follows by translation covariance.
\end{proof}
Theorem \ref{thm:distalsplit} establishes the Reeh-Schlieder property (and all the other consequences of the modular nuclearity condition discussed in Section \ref{sec:alg}) for double cones having a minimal ``relativistic size''. This size is measured by the length $s_{\min}$ and depends on the scattering function $S_2$ and the mass $m$. For example, if we consider a scattering function of the form \eqref{ex:s2} with $N=1$ and $\beta_1=\frac{i\pi}{4}$, one can derive the estimate $s_{\min}<l_C$, where $l_C$ is the Compton wavelength corresponding to the mass $m$.

Whereas the occurrence of a minimal localization length in theories describing quantum effects of gravity is expected for physical reasons, we conjecture that the minimal length $s_{\min}$ appearing here is an artifact of our estimates. This conjecture is supported by a second theorem, stated below, which improves the previous one under an additional assumption on the underlying scattering function.
\\
\\
The set $\SF_0$ of regular scattering functions can be divided into a ``Bosonic'' and a ``Fermionic'' class according to
\begin{align}\label{def:SF0pm}
\SF_0^\pm	&:=	\{S_2\in\SF_0\,:\,S_2(0)=\pm1\}\,,\quad& \SF_0&=\SF_0^+\cup\SF_0^-\,.
\end{align}
We emphasize that, independently of the scattering function, all the models under consideration describe Bosons in the sense that their scattering states are completely symmetric (see Section \ref{sect:s}). 

However, as will be shown below, there exist certain distinguished unitaries $Y^{\pm}$ mapping a model with $S_2\in\SF_0^\pm$ onto the Hilbert space $\Hil^\pm$ corresponding to the special model with the constant scattering function $S_2=\pm 1$.

In order to distinguish between the different scattering functions involved, we adopt the convention that the usual notations $z,\zd,D_n,P_n,\Hil_n,\Hil$ refer to the generic $S_2\in\SF_0$ under consideration. All objects corresponding to the constant scattering functions $S_2=\pm 1$ are tagged with an index ``$\pm$'', i.e. we write $z_\pm,\zd_\pm,D_n^\pm,P_n^\pm,\Hil_n^\pm,\Hil^\pm$.
\\
\\
In preparation for the construction of the unitaries $Y^\pm:\Hil\to\Hil^\pm$, note that each $S_2\in\mathcal{S}_0$ is analytic and nonvanishing in the strip $S(-\kappa(S_2),\kappa(S_2))$, since zeros and poles are related by $S_2(-\zeta)=S_2(\zeta)^{-1}$ (cf. \eqref{s2rel} and Definition \ref{def:S0}). So there exists an analytic function $\delta:S(-\kappa(S_2),\kappa(S_2))\to\Cl$ (the phase shift) such that
\begin{align}
  S_2(\zeta)
  &=
  S_2(0)\,e^{2i\delta(\zeta)},
  \qquad
  \zeta\in S(-\kappa(S_2),\kappa(S_2))\,.
\end{align}
Since $S_2$ has modulus one on the real line, $\delta$ takes real values on $\Rl$, and we fix it uniquely by the choice $\delta(0)=0$. Note that in view of $S_2(-\te)=\overline{S_2(\te)}$, $\te\in\Rl$, $\delta$ is odd.
\begin{lemma}\label{lemma:y}
  Let $S_2\in\SF_0^\pm$ and $\delta:S(-\kappa(S_2),\kappa(S_2))\longrightarrow\Cl$ be defined as above. Consider the functions
  \begin{align}\label{def-yn}
    Y_0^\pm=1,\qquad Y_1^\pm(\zeta)=1\,,\qquad
    Y_n^\pm (\bze)
    &:=
    \prod_{1\leq k < l \leq n} 
    \left(
      \pm
      e^{i\delta(\zeta_k-\zeta_l)}\right),\;n\geq 2, 
  \end{align}
  and the corresponding multiplication operators (denoted by the same
  symbol $Y_n^\pm$).
  \begin{enumerate}
  \item Viewed as an operator on $H^2(\Tu_n(\kappa(S_2)))$, $Y_n^\pm$ is a bounded map with operator norm $\|Y_n^\pm\|\leq\|S_2\|^{\,n/2}$.
  \item Viewed as an operator on $L^2(\Rl^n)$, $Y_n^\pm$ is a unitary intertwining the representations $D_n$ and $D_n^\pm$ of $\frS_n$, and hence mapping the $S_2$-symmetric subspace $\Hil_n\subset L^2(\Rl^n)$ onto the totally (anti-) symmetric subspace $\Hil_n^\pm\subset L^2(\Rl^n)$.
 \end{enumerate}
\end{lemma}
\begin{proof}
a) Since $\delta$ is analytic in $S(-\kappa(S_2),\kappa(S_2))$, so is the function $Y_n^\pm$ in the product domain  $S(-\frac{1}{2}\kappa(S_2),\frac{1}{2}\kappa(S_2))^{\times n}$. Depending only on differences $\zeta_k-\zeta_l$ of rapidities, $Y_n^\pm$ is also analytic in the tube $\Tu_n(\kappa(S_2))=S(-\frac{1}{2}\kappa(S_2),\frac{1}{2}\kappa(S_2))^{\times n}+ i\bla_0$. In view of \eqref{bnd-Srho}, it follows that
  \begin{align}\label{y-bound}
    \left|
      Y_n^\pm(\bze)
    \right|
  &\leq
  \|S_2\|^{n/2}
  \,,\qquad \bze\in\Tu_n(\kappa(S_2))\,.
\end{align}
Hence $Y_n^\pm$ maps $H^2(\Tu_n(\kappa(S_2)))$ into itself, and the bound $\bno{Y_n^\pm F} \leq \|S_2\|^{\frac{n}{2}}\bno{F}$, $F\in H^2(\Tu_n(\kappa(S_2)))$, proves a).

b) Considered as a multiplication operator on $L^2(\Rl^n)$, $Y_n^\pm$ multiplies with a phase and is hence unitary. Let $\tau_j\in\frS_n$ denote the transposition exchanging $j$ and $j+1$, $j\in\{1,...,n\}$, and pick arbitrary $f_n \in L^2(\Rl^n)$, $\bte\in\Rl^n$. 
 \begin{align*}
   (D_n^\pm(\tau_j)Y_n^\pm f_n)&(\bte)
   =
   e^{i\delta(\te_{j+1}-\te_j)}\prod_{\genfrac{}{}{0pt}{}{1\leq k < l \leq n}{(k,l)\neq (j,j+1)}}\!\!
   \left(\pm e^{i\delta(\te_k-\te_l)}\right)
   f_n(\te_1,..,\te_{j+1},\te_j,..,\te_n)\\
   &=
   \prod_{1\leq k < l \leq n}
   \left(\pm e^{i\delta(\te_k-\te_l)}\right)
   \cdot
   S_2(\te_{j+1}-\te_j)
   f_n(\te_1,..,\te_{j+1},\te_j,..,\te_n)\\
   &=
   (Y_n^\pm D_n(\tau_j)f_n)(\bte)
  \end{align*}
As the transpositions $\tau_j$ generate $\frS_n$, this calculation shows that $Y_n^\pm $ intertwines $D_n^\pm$ and $D_n$. In particular, $Y_n^\pm$ restricts to a unitary mapping $\Hil_n$ onto $\Hil_n^\pm$.
\end{proof}

The operator
\begin{align}
Y^\pm := \bigoplus_{n=0}^\infty Y_n^\pm : \Hil\longrightarrow\Hil^\pm
\end{align}
will be used to improve the estimate on $\|\Xi_n(s)\|_1$ underlying Theorem \ref{thm:distalsplit}. In a model theory with scattering function $S_2\in\SF_0^\pm$, we consider the maps 
\begin{align*}
  \Xi_n^\pm(s)
	&:=
	Y_n^\pm\Xi_n(s) \,:\,  \M \longrightarrow \Hil_n^\pm
  \,,\qquad \Xi^\pm(s):=Y^\pm\Xi(s)\,.
\end{align*}
Since $Y^\pm : \Hil\to\Hil^\pm$ is unitary, $\Xi(s)$ is nuclear if and only if $\Xi^\pm(s)$ is, and in this case $\|\Xi(s)\|_1=\|\Xi^\pm(s)\|_1$. Moreover, as $Y_n^\pm$ acts by multiplication with a function depending only on differences of rapidities, this operator commutes with the translation $U(\us)$ and the modular operator $\Delta$, i.e.
\begin{align*}
  \Xi_n^\pm(s)A 
  &=
  \Delta^{1/4}U(\tfrac{1}{2}\us)Y_n^\pm
  (A(\tfrac{1}{2}s)\Omega)_n
  \;=:\;
  \left(\Delta_n^\pm(s,\kappa) \; Y_n^\pm \Sigma_n(s,\kappa)\right)A
  \,.
\end{align*}
Here $\Sigma_n(s,\kappa)$ is defined as in \eqref{def:Sigma_n} and $\Delta^\pm_n(s,\kappa)$ acts as $\Delta_n(s,\kappa)$
\eqref{def:Delta_n}, but is now considered as a map from the subspace $H^2_\pm(\Tu_n(\kappa))\subset H^2(\Tu_n(\kappa))$, consisting of the totally (anti-) symmetric functions in $H^2(\Tu_n(\kappa))$, to $\Hil_n^\pm$.

Lemma \ref{cor:hardy} and Lemma \ref{lemma:y} a) imply that $Y_n^\pm\Sigma_n(s,\kappa)$ is a bounded linear map from $\M$ to $H^2_\pm(\Tu_n(\kappa))$, $\kappa\in(0,\kappa(S_2))$, with norm
\begin{align}\label{sigmaprime}
  \|Y_n^\pm \Sigma_n(s,\kappa)\|
  &\leq
	\left(  \|S_2\|^{1/2}\cdot\sigma(s,\kappa)\right)^n
  \,.
\end{align}
In the case $S_2\in\SF_0^-$, the Pauli principle effectively reduces the size of the image of $\Delta_n^-(s,\kappa)$, which results in an improved estimate on $\|\Xi(s)\|_1$, implying the following theorem. In the case $S_2(0)=+1$, the Pauli principle does not apply and the subsequent argument cannot be used to obtain nuclearity for arbitrarily small splitting distances. It should be mentioned, however, that the scattering functions of all models known from Lagrangian formulations belong to the class $\SF_0^-$ \cite{kb1}.
\begin{theorem}{\bf(Proof of the modular nuclearity condition)}\label{thm:-1nuclearity}\\
In a model theory with scattering function $S_2\in\SF_0^-$ \eqref{def:SF0pm}, the maps $\Xi(s)$ are nuclear for every splitting distance $s>0$.

In particular, in these models there exist observables localized in arbitrarily small open regions  $\OO\subset\Rl^2$, and the Reeh-Schlieder property holds without restriction.
\end{theorem}
\begin{proof}
Proceeding along the same lines as in the proof of Lemma \ref{lemma:delta}, we infer that $\Delta_n^-(s,\kappa)$ is nuclear and can be represented as in \eqref{delta-eqn}. With the notations used there, $\beps=(\eps_1,...,\eps_n)$, $\eps_k=\pm 1$, there holds for $F^-\in H^2_-(\Tu_n(\kappa))$
\begin{align}\label{xi-t}
 \Delta_n^-(s,\kappa)F^-
 &=
 2^{-n}\sum_{\sbeps} \eps_1\cdots\eps_n (T_{s,\eps_1\kappa}\otimes ... \otimes T_{s,\eps_n\kappa}) 
    F^-_{\sbla_0-\frac{\kappa}{2}\sbeps}\;.
\end{align}
Choosing an orthonormal basis $\{\psi_k\}_k$ of $L^2(\Rl)$, the vectors
\begin{align}\label{ONB}
  \Psi_\sbk^-
  &:=
  \zd_-(\psi_{k_1})\cdots\zd_-(\psi_{k_n})\Omega
  \;=\;
  \sqrt{n!}\,P_n^-(\psi_{k_1}\otimes ... \otimes \psi_{k_n})
	\nonumber
	\\
  &=
	\frac{1}{\sqrt{n!}}\sum_{\rho\in\frS_n}{\rm sign}(\rho)\,\psi_{\rho(k_1)}\otimes...\otimes\psi_{\rho(k_n)}
\end{align}
form an orthonormal basis of $\Hil^-_n$ if $\bk=(k_1,...,k_n)$ varies over $k_1<k_2<...<k_n$, $k_1,...,k_n\in\N$, as a consequence of the Pauli principle.

Expanding the right hand side of \eqref{xi-t} in this basis, we find
\begin{align*}
\Delta_n^-(s,\kappa)F^-
&=
2^{-n}\sum_{\sbeps}\eps_1\cdots\eps_n\!\!
\sum_{k_1<...<k_n}
\langle \Psi_\sbk^-,(T_{s,\eps_1\kappa}\otimes .. \otimes T_{s,\eps_n\kappa})F^-_{\sbla_0-\frac{\kappa}{2}\sbeps}\rangle
\,\Psi_\sbk^-
\\
&=
\frac{2^{-n}}{\sqrt{n!}}\sum_{\sbeps}\eps_1\cdots\eps_n\sum_{\rho\in\frS_n}{\rm sign}(\rho)\times\\
&\qquad\times\sum_{k_1<...<k_n}
\langle T_{s,\eps_1\kappa}^*\psi_{\rho(k_1)}\otimes .. \otimes T_{s,\eps_n\kappa}^*\psi_{\rho(k_n)},\,
F^-_{\sbla_0-\frac{\kappa}{2}\sbeps}\rangle
\,\Psi_\sbk^-
\,.
\end{align*}
This is an example of a nuclear decomposition \eqref{nuc-decc} of $\Delta^-_n(s,\kappa)$, with the functionals $\eta_k$ from Definition \ref{def:nucmap} being given by
\begin{equation}\label{etaexp}
 \eta^\psi_{\sbeps,\rho,\sbk}(F^-) := \frac{\eps_1\cdots\eps_n\,{\rm sign}(\rho)}{2^n\sqrt{n!}}
\langle T_{s,\eps_1\kappa}^*\psi_{\rho(k_1)}\otimes ... \otimes T_{s,\eps_n\kappa}^*\psi_{\rho(k_n)},\,
F^-_{\sbla_0-\frac{\kappa}{2}\sbeps}\rangle
\,.
\end{equation}
To obtain a good bound on $\|\eta^\psi_{\sbeps,\rho,\sbk}\|$, we have to choose the basis $\{\psi_k\}_k$ in an appropriate way. Consider the positive operator $\hat{T}_{s,\kappa} := (|T_{s,\kappa}^*|^2 + |T_{s,-\kappa}^*|^2)^{1/2}$, which is of trace class on $L^2(\Rl)$ and satisfies $\|\hat{T}_{s,\kappa}\|_1 \leq 2\,\|T_{s,\kappa}\|_1$ \cite{kosaki}. We choose $\{\psi_k\}_k$ as normalized  eigenvectors of $\hat{T}_{s,\kappa}$, with eigenvalues $t_k\geq 0$.

Noting $\|T_{s,\pm\kappa}^*\psi_{k_j}\|\leq \|\hat{T}_{s,\pm\kappa}\psi_{k_j}\|=t_{k_j}$ and $\|F^-_{\sbla_0-\frac{\kappa}{2}\sbeps}\|\leq\bno{F^-}$, we can estimate \eqref{etaexp} according to $\|\eta^\psi_{\sbeps,\rho,\sbk}(F^-)\|\leq t_{k_1}\cdots t_{k_n}/(2^n\sqrt{n!})\cdot\bno{F^-}$. Since $\|\Psi_\sbk^-\|=1$ and $\sum_{\sbeps,\rho}1=2^n\cdot n!$, this yields
\begin{align*}
\|\Delta_n^-(s,\kappa)\|_1
\leq
\sqrt{n!}
\sum_{k_1<...<k_n}t_{k_1}\cdots t_{k_n}
\leq
\frac{1}{\sqrt{n!}}
\sum_{k_1,...,k_n=1}^\infty t_{k_1}\cdots t_{k_n}
=
\frac{\|T_{s,\kappa}\|_1^{\,n}}{\sqrt{n!}}
\;.
\end{align*}
In view of the bound \eqref{sigmaprime} on $Y_n^-\Sigma_n(s,\kappa)$, we arrive at the following estimate for the nuclear norm of $\Xi^-(s)=\sum_{n=0}^\infty \Delta^-_n(s,\kappa) Y_n^-\Sigma_n(s,\kappa)$,
\begin{align}
\|\Xi^-(s)\|_1
&\leq
\sum_{n=0}^\infty \frac{\left(\sigma(s,\kappa)\,\|S_2\|^{1/2}\,\|T_{s,\kappa}\|_1\right)^n}{\sqrt{n!}}<\infty\,.
\end{align}
This series converges for arbitrary values of 
$\sigma(s,\kappa)\|S_2\|^{1/2}\,\|T_{s,\kappa}\|_1$, i.e. for arbitrary splitting distances $s>0$.
\end{proof}

\section{Collision States and Reconstruction of the S-Matrix}\label{sect:s}

The theorems of the preceding section establish the existence of a class of quantum field theories. In this section, we investigate the collision states of these models and prove that they provide the solution of the inverse scattering problem for the considered class of S-matrices. More precisely, we will show that the function $S_2$, which entered as a parameter into the construction, is related to the S-matrix of the model as in \eqref{S-mult} (Theorem \ref{thm:S-matrix}). Moreover, we will find explicit formulae for $n$-particle scattering states and give a proof of asymptotic completeness (Proposition \ref{prop:ac}).

To compute $n$-particle collision states, it is sufficient to restrict to the family $\SF_0$ of regular  scattering functions (Definition \ref{def:S0}), as Theorem \ref{thm:distalsplit} ensures that in this case there exist compactly localized observables satisfying the Reeh-Schlieder property, at least in double cones above some minimal size. Since any number of double cones of {\em any} size can be spacelike separated by translation, it is possible to apply the usual methods of collision theory in this class of theories -- localization with arbitrarily high precision is not needed.

The method to be used for the calculation of the S-matrix is Haag-Ruelle scattering theory \cite[Ch. 5]{araki} in the same form as in \cite{BBS}, where scattering properties of polarization-free generators have been analyzed. As usual in this approach, we consider quasilocal operators of the form, $A\in\A(\OO)$, 
\begin{align}\label{a-quasilocal}
A(f_t)
&=
\int d^2x\,
f_t(x)A(x)\,,\qquad A(x)=U(x)A\,U(x)^{-1}\,,
\end{align}
The functions $f_t$, $t\in \Rl$, are defined in terms of momentum space wavefunctions $\fti$ by
\begin{align}\label{def:ft}
f_t(x)
&:=
\frac{1}{2\pi}
\int d^2p\,
\fti(p_0,p_1)\,e^{i(p_0-\omega_p)t}\,e^{-ip\cdot x}\;,\qquad \omega_p := \big(m^2+p_1^2\big)^{1/2}\,.
\end{align}
Here $\fti$ is taken to be a Schwartz test function, such that the integral \eqref{a-quasilocal} converges in operator norm.

For the construction of collision states, the asymptotic properties as $t\to\pm\infty$ of these functions are important. We introduce the {\em velocity support} of $f$ as
\begin{align}
\VV(f)
&:=
\big\{(1,p_1\cdot\omega_p^{-1})\,:\,(p_0,p_1)\in\supp\fti\,\big\}\,,\qquad\omega_p:=(p_1^2+m^2)^{1/2}\,.
\label{velsup}
\end{align}
Recall that the support of $f_t$ is essentially contained in $t\,\VV(f)$ for asymptotic times $t$ \cite{hepp}. More precisely, let $\chi$ be a smooth function which is equal to 1 on $\VV(f)$ and vanishes in the complement of a slightly larger region. Then the asymptotically dominant part of $f_t$ is $\fhat_t(x):=\chi(x/t)f_t(x)$, and for any $N\in\N$, the difference $|t|^N(f_t-\fhat_t)$ converges to zero in the topology of $\Ss(\Rl^2)$ as $t\to\pm\infty$ (\cite{hepp}, see also \cite[Cor. to Thm. XI.14]{SimonReed3}). We also adopt the notation from \cite{BBS} to write $f\prec g$ if $\VV(g)-\VV(f)\subset \{0\}\times(0,\infty)$.

This notation will be used for single particle wavefunctions as well: Given smooth, compactly supported $\te\mapsto\psi_1(\te),\te\mapsto\psi_2(\te)$, we write $\psi_1\prec\psi_2$ if $\supp\psi_2-\supp\psi_1\subset(0,\infty)$. It is straightforward to show that in this situation, there exist testfunctions $f_1,f_2\in\Ss(\Rl^2)$ such that $f_1^+=\psi_1$, $f_2^+=\psi_2$, and $f_1\prec f_2$ in the previously defined sense.
\\
\\
If the support of $\fti$ is concentrated around a point $(\omega_p,p_1)$ on the upper mass shell and does not intersect the energy momentum spectrum elsewhere, $A(f_t)\Om\in\Hil_1$ is a single particle state which does not depend on the time parameter $t$. Furthermore, there exist the following (strong) limits
\begin{align}
\lim_{t\to\pm\infty}A(f_t)\Psi	=	A(f)_{\genfrac{}{}{0pt}{}{\rm out}{\rm in}}\Psi
\,,\qquad
\lim_{t\to\pm\infty}A(f_t)^*\Psi	=	A(f)_{\genfrac{}{}{0pt}{}{\rm out}{\rm in}}^*\Psi\,,
\end{align}
to the asymptotic creation and annihilation operators $A(f)_{\rm out/in}$ and $A(f)_{\rm out/in}^*$, respectively.

These limits are known to hold for all scattering states $\Psi$ of compact energy momentum support, in particular, for all single particle states of the form $\phi(f)\Om=f^+$, where $f^+$ has compact support \cite{BBS}.

The creation and annihilation operators ${A(f)\ex}^{(*)}$, $\rm ex=in/out$, are related to the Zamolodchikov operators $\zd_+$, $z_+$ with the constant scattering function $S_2=1$, acting on the totally symmetric Bose Fock $\Hil^+$ space over $\Hil$. This relation is implemented by the M{\o}ller operators $V\ex : \Hil^+\to\Hil$,
\begin{align}
A(f)\ex		&=	V\ex\,\zd_+\big(A(f)\Om\big)\,{V\ex}^*,\quad 
{A(f)\ex}^*	=	V\ex\, z_+\big(\overline{A(f)\Om}\big)\, {V\ex}^*\,.
\end{align}
Having recalled these basic facts of scattering theory, we now fix a regular scattering function and compute $n$-particle collision states in the corresponding model theory. Using the standard notation for scattering states, we find the following Lemma.
\begin{lemma}{\bf(Calculation of $n$-particle collision states)}\label{lemma-scatter}\\
Consider testfunctions $\fti_1,...,\fti_n\in\Ss(\Rl^2)$ having pairwise disjoint compact supports concentrated around points on the upper mass shell such that $f_1\prec...\prec f_n$. Then
\begin{align}
(f_1^+ \times ... \times f_n^+)\oout
&=
\phi(f_1)\cdots\phi(f_n)\Om
=
\sqrt{n!}\,P_n(f_1^+\otimes...\otimes f_n^+)
\,,\label{n-out}\\
(f_1^+\times ... \times f_n^+)\iin
&=
\phi(f_n)\cdots\phi(f_1)\Om
=
\sqrt{n!}\,P_n(f_n^+\otimes...\otimes f_1^+)
\,.\label{n-in}
\end{align}
\end{lemma}
\begin{proof}
Since the supports of the $\fti_k$ do not intersect the lower mass shell, the annihilation parts of the fields $\phi(f_k)$ vanish, $\phi(f_k)=\zd(f_k^+)$. So the second identity in \eqref{n-out} and \eqref{n-in} follows from \eqref{def:zzd}.
 
The proof of the first identity in \eqref{n-out} and \eqref{n-in} is based on induction in the particle number $n$. For $n=1$, we have
\begin{align}
\phi(f_1)\Om	=	f_1^+	=	(f_1^+)\oout	=	(f_1^+)\iin
\,,
\end{align}
since $f_1^+$ is a single particle state. For the step from $n$ to $(n+1)$ particles, consider operators $A_1,...,A_n\in\A(\OO)$ localized in a double cone $\OO$ large enough for $\Om$ to be cyclic for $\A(\OO)$. We want to establish commutation relations between $\phi(f)$ and the creation operators $A_k(g_k)\oout$, where $f\prec g_1\prec ... \prec g_n$ and the test functions $f, g_1,...,g_n$ have the same support properties as the $f_1,...,f_n$. As the support of $\fti$ intersects the energy momentum spectrum only in the upper mass shell, it readily follows from the definitions \eqref{def:phi} of $f^\pm$ and \eqref{def:ft} of $f_t$, that $f_t^+=f^+$, $f_t^-=0$, $t\in\Rl$. Since $\fhat_t-f_t$ converges to zero in $\Ss(\Rl^2)$ for $t\to\infty$, and since $f\mapsto\phi(f)\Psi$, $\Psi\in\DD$, is a vector valued tempered distribution, this implies
\begin{align}\label{phi-limit}
\phi(f)\Psi	=	\phi(f_t)\Psi	=	\lim_{t\to\infty}\phi(\fhat_t)\Psi
\,,\qquad
\Psi\in\DD\,.
\end{align}
Since $\|A(\ghat_{k,t})-A(g_{k,t})\|\leq \|A\|\|\ghat_{k,t}-g_{k,t}\|_1\to 0$ for $t\to\infty$, we can use the strong convergence $A_k(g_{k,t})\to A_k(g_k)\oout$ and the hermiticity of $\phi$ to obtain, $\Psi\in\DD$,
\begin{align*}
\langle\phi(f)^*\Psi,(A_1(g_1)\Om\times .. \times A_n(g_n)\Om)\oout\rangle
&=
\lim_{t\to\infty}\langle\phi(\fhat_t)^*\Psi,A_1(\ghat_{1,t})\cdots A_n(\ghat_{n,t})\Om\rangle
.
\end{align*}
For large $t$, the functions $\hat{f}_t$ and $\hat{g}_{k,t}$ have supports in small neighborhoods of $t\,\VV(f)$ and $t\,\VV(g_k)$, respectively. Hence $\phi(\fhat_t)^*$ is localized in a wedge $W_L^{(t)}$ slightly larger than $W_L+t\,\VV(f)$, and $A_k(\hat{g}_{k,t})$ is localized in a neighborhood of $\OO+t\,\VV(g_k)$. For large enough $t>0$, these regions are spacelike separated since $f\prec g_k$. As $\phi(\fhat_t)^*$ is affiliated with $\A(W_L^{(t)})$, it follows that this operator commutes with $A_k(\ghat_{k,t})$, $k=1,...,n$. Thus
\begin{align*}
\langle\phi(f)^*\Psi,(A_1(g_1)\Om\times ... \times A_n(g_n)\Om)\oout\rangle
&=
\lim_{t\to\infty}\langle\Psi\,,A_1(\ghat_{1,t})\cdots A_n(\ghat_{n,t})\phi(\fhat_t)\Om\rangle
\\
&=
\lim_{t\to\infty}\langle\Psi\,,A_1(\ghat_{1,t})\cdots A_n(\ghat_{n,t})\, \fhat_t^+\rangle
\,.
\end{align*}
A straightforward estimate yields $\|A_1(\ghat_{1,t})\cdots A_n(\ghat_{n,t})\|\leq c\,t^{2n}$ with a constant $c>0$. But since $t^{2n}(\fhat_t-f_t)$ converges to zero in the topology of $\Ss(\Rl^2)$, it follows that also  $t^{2n}\|\fhat_t^+-f^+\|_2\to 0$. So we may replace $\fhat_t^+$ in the above equation by $f^+$, and use the strong convergence $A_k(\ghat_{k,t})\to A_k(g_k)\oout$ on this single particle state to conclude
\begin{align*}
\langle\phi(f)^*\Psi,(A_1(g_1)\Om\times ... \times A_n(g_n)&\Om)\oout\rangle
=
\langle \Psi, A_1(g_1)\oout \cdots A_n(g_n)\oout f^+\rangle
\\
&=
\langle \Psi, (A_1(g_1)\Om\times ... \times A_n(g_n)\Om\times f^+)\oout\rangle
\\
&=
\langle \Psi, (f^+\times A_1(g_1)\Om\times ... \times A_n(g_n)\Om)\oout\rangle
\,,
\end{align*}
where in the last step we used the Bose symmetry of the scattering states.

In view of the Reeh-Schlieder property of $\A(\OO)$, we can approximate the single particle state $f_k^+$ by $A_k(g_k)\Om$. Given any $\eps>0$, there exist local operators $A_1,...,A_n\in\A(\OO)$ and functions $g_1,...,g_n$, with $g_k$ having support in an arbitrarily small neighborhood of the support of $f_k$, such that $\|f_k^+-A_k(g_k)\Om\|<\eps$. As the left and right hand side of the above equation are continuous in the $A_k(g_k)\Om$, this implies
\begin{align}
\langle\phi(f)^*\Psi,(f_1^+\times ... \times f_n^+)\oout\rangle
&=
\langle\Psi\,,(f^+\times f_1^+\times ... \times f_n^+)\oout\rangle
\,.
\end{align}
Since $\Psi\in\DD$ was arbitrary and $\DD\subset\Hil$ is dense, we can use the induction hypothesis and obtain
\begin{align*}
\phi(f)\phi(f_1)\cdots \phi(f_n)\Om
&=
\phi(f)(f_1^+\times ... \times f_n^+)\oout
=
(f^+\times f_1^+\times ... \times f_n^+)\oout
\,,
\end{align*}
proving \eqref{n-out}.

For incoming $n$-particle states, the order of the velocity supports of $f_1,...,f_n$ has to be reversed, since $W_L+t\,\VV(f_1)$ is spacelike separated from $\OO+t\,\VV(f_k)$ for $t\to-\infty$ if $f\succ f_k$. Apart from this modification, the same argument can be used to derive formula \eqref{n-in}.
\end{proof}

Given smooth, compactly supported single particle functions $\psi_1,...,\psi_n\in\Hil_1$ with supports ordered according to $\psi_1\prec...\prec\psi_n$, there exist testfunctions $f_1,...,f_n\in\Ss(\Rl^2)$ such that $f_k^+=\psi_k$, $f_k^-=0$, $k=1,...,n$, and $f_1\prec..\prec f_n$. Hence for these $\psi_k$,
 \begin{align}
(\psi_1\times ... \times \psi_n)\oout
&=
\sqrt{n!}\,P_n(\psi_1\otimes ... \otimes \psi_n)
\,,\qquad \psi_1\prec...\prec\psi_n\,,\label{n-out2}
\\
(\psi_1\times ... \times \psi_n)\iin
&=
\sqrt{n!}\,P_n(\psi_n\otimes ... \otimes \psi_1)
\,,\qquad \psi_1\prec...\prec\psi_n\,.\label{n-in2}
\end{align}
In terms of improper $n$-particle states with sharp rapidities, we have thus shown that
\begin{subequations}\label{z-in-out}
\begin{align}
\zd(\te_1)\cdots\zd(\te_n)\Om	&=	|\,\te_1,...,\te_n\rangle\oout
\,,&\quad
&\te_1<...<\te_n
\,,\\
\zd(\te_1)\cdots\zd(\te_n)\Om	&=	|\,\te_1,...,\te_n\rangle\iin
\,,&\quad
&\te_1>...>\te_n
\,,
\end{align}
\end{subequations}
are asymptotic collision states in the sense of the Haag-Ruelle scattering theory.

The identification of incoming and outgoing $n$-particle states with $n$-fold products of such creation operators acting on the vacuum, arranged in order of decreasing, respectively increasing, rapidities, is one of the basic assumptions in the framework of the form factor program. In fact, it has motivated the very definition of the Zamolodchikov-Faddeev algebra \cite{ZZ}. It is therefore gratifying that with the help of the approach presented here, the heuristic picture underlying the relations of this algebra can be rigorously justified. 
\\\\
The outgoing and incoming scattering states (\ref{n-out2}, \ref{n-in2}) form total sets in the Hilbert space $\Hil$. To prove this, note that the functions $\bte\mapsto\prod_{k=1}^n\psi_k(\te_k)$ form a total set in the space $L^2(E_n)$ of all square integrable functions on the simplex $E_n:=\{(\te_1,...,\te_n)\in\Rl^n\,:\,\te_1\leq...\leq\te_n\}$ when the $\psi_k$ are varied within the limitations specified above. But the $S_2$-symmetrization $P_n$ is a linear and continuous map from $L^2(E_n)$ to $\Hil_n$, with dense range. Hence the totality of the constructed outgoing $n$-particle collision states in $\Hil_n$ follows. 

Analogously, one can show that also the incoming $n$-particle states form a total set in $\Hil_n$. Taking linear combinations of states of different particle number, it also follows that the spaces $\Hil_{\rm out}$ and $\Hil_{\rm in}$ spanned by all outgoing and incoming scattering states are dense in $\Hil$. So we arrive at the following proposition.

\begin{proposition}{\bf(Asymptotic completeness)}\label{prop:ac}\\
All model theories with regular scattering functions are asymptotically complete.
\end{proposition}
This result seems to be the first proof of asymptotic completeness in an interacting relativistic quantum field theory \cite{BuSu-scatter}.
\\\\
We finish this section by computing the M{\o}ller operators $V\iin$, $V\oout$ and the S-matrix $S$. The asymptotic states span the Bosonic Fock space $\Hil^+=\bigoplus_{n=0}^\infty \Hil_n^+$ over $\Hil_1=L^2(\Rl)$. Denoting the orthogonal projection onto $\Hil_n^+$ by $P_n^+$, we infer from the form (\ref{n-out2},\ref{n-in2}) of the collision states that the M{\o}ller operators are given by
\begin{align}
V\oout P_n^+(\psi_1\otimes ...\otimes \psi_n)	&=	P_n(\psi_1\otimes...\otimes\psi_n)
\,,\qquad \psi_1\prec...\prec\psi_n\,,\label{vnout}
\\
V\iin P_n^+(\psi_n\otimes ...\otimes \psi_1)	&=	P_n(\psi_n\otimes...\otimes\psi_1)\label{vnin}
\,,\qquad \psi_1\prec...\prec\psi_n\,.
\end{align}
In view of the ordering of the supports of the $\psi_k$, these equations determine two well-defined linear operators $V_{\rm in/out}$ with dense domains and ranges, and since  
\begin{equation}
 \|P_n^+(\psi_1\otimes...\otimes\psi_n)\|=n!^{-1/2}\|\psi_1\|\cdots\|\psi_n\|=\|P_n(\psi_1\otimes...\otimes\psi_n)\|\,,
\end{equation}
$V\iin$ and $V\oout$ continue to unitaries mapping $\Hil^+$ onto $\Hil$. The S-matrix is the product of the M{\o}ller operators,
\begin{equation}\label{def:S}
 S:={V\oout}^*V\iin:\Hil^+\to\Hil^+\,.
\end{equation}
\begin{theorem}{\bf(Calculation of the S-matrix)}\label{thm:S-matrix}\\
The model with scattering function $S_2\in\SF_0$ solves the inverse scattering problem for the corresponding S-matrix, i.e. its scattering operator \eqref{def:S} is, $\Psi^+\in\Hil^+$,
\begin{align}\label{S-form}
(S\Psi^+)_n(\te_1,...,\te_n)
=
\prod_{1\leq l<k\leq n}S_2(|\te_l-\te_k|)
\cdot \Psi^+_n(\te_1,...,\te_n)
\,.
\end{align}
\end{theorem}
\begin{proof}
Recall that the $S_2$-symmetrization operator $P_n$ has the form \eqref{def:Dn}
\begin{align}
(P_n \Psi_n)(\te_1,...,\te_n)
&=
\frac{1}{n!}\sum_{\rho\in\frS_n}
S_n^\rho(\te_1,...,\te_n)\cdot\Psi_n(\te_{\rho(1)},...,\te_{\rho(n)})\,,\\
S_n^\rho(\te_1,...,\te_n)
&=
\prod_{\genfrac{}{}{0pt}{}{1\leq l < k \leq n}{\rho(l) > \rho(k)}} S_2(\te_{\rho(l)}-\te_{\rho(k)})\,.\label{snexp}
\end{align}
Consider $\psi_1,...,\psi_n\in C_0^\infty(\Rl)$, $\psi_1\prec...\prec\psi_n$ and a point $\bte\in\Rl^n$ such that  $\te_{\pi(1)}<...<\te_{\pi(n)}$ for some permutation $\pi\in\frS_n$. In this situation, there holds
\begin{equation}
 (P_n(\psi_1\otimes...\otimes\psi_n))(\bte) = \frac{1}{n!}\,S_n^\pi(\bte)\cdot\psi_1(\te_{\pi(1)})\cdots\psi_n(\te_{\pi(n)})\,,
\end{equation}
and $V\oout$ \eqref{vnout} is seen to act on $n$-particle states by multiplication with the function $\bte\mapsto\{S_n^\pi(\bte)\,:\,\te_{\pi(1)}\leq...\leq\te_{\pi(n)}\}$. Similarly, $V\iin$ \eqref{vnin} acts by multiplication with $\bte\mapsto\{S_n^{\pi\iota}(\bte)\,:\,\te_{\pi(1)}\leq...\leq\te_{\pi(n)}\}$, where $\iota\in\frS_n$ is the total inversion permutation, $\iota(k):=n-k+1$. This implies that the $n$-particle S-matrix is the multiplication operator
\begin{eqnarray}
 (S\Psi^+)_n(\bte)	&=& \hat{S}_n(\bte)\cdot\Psi_n^+(\bte)\,,\qquad\;\;\Psi^+\in\Hil^+\,,\\
\hat{S}_n(\bte)     &:=&\{S_n^\pi(\bte)^{-1}S_n^{\pi\iota}(\bte)\,:\,\te_{\pi(1)}\leq...\leq\te_{\pi(n)}\}\,.
\end{eqnarray}
Since $D_n$ \eqref{def:Dn} is a representation of $\frS_n$, there holds
\begin{equation}
 S_n^{\pi\iota}(\te_1,...,\te_n)=S_n^\pi(\te_1,...,\te_n)S_n^{\,\iota}(\te_{\pi(1)},...,\te_{\pi(n)})\,,\qquad \te_1,...,\te_n\in\Rl\,.
\end{equation}
Hence, for $\te_{\pi(1)}<...<\te_{\pi(n)}$,
\begin{align}
 \hat{S}_n(\te_1,...,\te_n)	&=	S_n^{\,\iota}(\te_{\pi(1)},...,\te_{\pi(n)})
= \prod_{1\leq l<k\leq n}S_2(\te_{\pi\iota(l)}-\te_{\pi\iota(k)})\\
&=\prod_{1\leq l<k\leq n}S_2(|\te_{\iota\pi(l)}-\te_{\iota\pi(k)}|)
=\prod_{1\leq l<k\leq n}S_2(|\te_l-\te_k|)\,.
\end{align}
As all reference to the permutation $\pi$ has been eliminated, this formula is valid for arbitrary $\te_1,...,\te_n\in\Rl$, and finishes the proof of the claimed expression \eqref{S-form} for the S-matrix.
\end{proof}

\section{Conclusions}

In the present article, the construction of a large class of quantum field theories with factorizing S-matrices has been completed. The starting point of this construction is a pair of wedge-local quantum fields associated with a given S-matrix $S$, and the observation that the structure of the local observables corresponding to $S$ are fixed by commutation relations with these fields. By employing operator-algebraic techniques, basic problems such as the existence of models with a prescribed S-matrix were solved without having to specify explicit formulae for local interacting quantum fields.

It is interesting to notice that, at least in the class of models considered here, the rather abstract modular nuclearity condition needed to prove the existence of local observables amounts to very explicit conditions of analyticity and boundedness properties of matrix elements of observables localized in wedges. So these form factors play an important role also in the construction presented here, although in a manner quite different from their use in the form factor program.

For a complete understanding of these models, both, the algebraic approach presented here and the form factor program, are relevant. Structural properties like asymptotic completeness (which enters into the form factor program as an assumption) can be more conveniently analyzed in the algebraic framework. Furthermore, it is possible to discuss large classes of models at the same time in this approach. In comparison, the form factor program is better suited for deriving approximate formulae for local quantities such as $n$-point Wightman functions. Although the convergence of the form factor expansion is not under control yet, one might speculate that this situation can be improved in view of the now established existence theorem, just as the heuristic motivation of the relations of the Zamolodchikov-Faddeev algebra were rigorously justified in Haag-Ruelle scattering theory. 

In addition to properties of the scattering states, also something about the thermodynamics of models with a factorizing S-matrix can be learned from our analysis. By a slight generalization of our arguments, and following the reasoning in \cite{BDL2}, it can be shown that the maps
$\Theta_\beta(s) : \A(\OO_s) \to \Hil$, $\Theta_\beta(s)A := e^{-\beta
  H}A\Omega$, where $H$ denotes the Hamiltonian and $\OO_s=W_R\cap(W_L+(0,s))$, $s>0$, are nuclear if $\Xi(s)$ is. An estimate on the nuclear norms $\|\Theta_\beta(s)\|_1$ can be calculated. As the quantity $\|\Theta_\beta(s)\|_1$ is to be interpreted as the partition function of the restriction of the considered theory to the ``relativistic box'' $\OO_s$ at inverse temperature $\beta$
\cite{BuWi}, such estimates provide information about gross thermodynamical properties of the system.
\\
\\
In the present paper, we restricted ourselves to models describing a single species of neutral, scalar particles. There also exist many integrable quantum field theories with richer particle spectra, containing bound states and solitons. The generalization of the construction procedure presented here to this larger class of models is currently under investigation\footnote{H.~Grosse and G.~Lechner, work in progress.}. Before a generalization to models with bound states can be realized, one probably needs to develop an operator-algebraic understanding of the singularity structure of the corresponding S-matrices \cite{abdalla,kb2}, just as the crossing symmetry of factorizing S-matrices is now known to be linked to the wedge-locality of its associated polarization-free generators \cite{sc}.

Besides these more specific aspects of models with factorizing S-matrices, we note in conclusion that the general idea of constructing interacting model theories by first considering nets of wedge algebras and then analyzing their relative commutants is applicable to higher-dimensional spacetimes as well. However, the modular nuclearity condition cannot be satisfied if the spacetime dimension is
larger than two. Finding an adequate condition, applicable in
physical spacetime and ensuring the non-triviality of intersections of wedge algebras, might therefore lead to considerable progress in the construction of interacting quantum field theories.

\begin{acknowledgements}
Since this article is the result of a rather long investigation, I have reason to thank many people and institutions. Many discussions with my PhD advisor D. Buchholz have been important for this work. Regarding the theory of complex analysis, I had the opportunity to learn a lot from H.-J. Borchers and J. Bros. G. Garrigos pointed out reference \cite{StW} to me, and M. Karowski, F.~A. Smirnov and A. Fring informed me about the status of the form factor program. I also benefitted from several conversations with B. Schroer, J. Mund, M. M\"uger, K.-H. Rehren and R. Verch during different stages of this work. I wish to thank them all.

Financial support by the Deutsche Forschungsgemeinschaft DFG, and travel grants by the universities of Gainesville and S\~ao Paulo, the DFG, the Daniel Iagolnitzer foundation and the Oberwolfach Institute are thankfully acknowledged. Last but not least, my thanks go to J.~Yngvason for inviting me to the Erwin Schr\"odinger Institute in Vienna, where the final version of this article was written up.
\end{acknowledgements}

\appendix
\section{Proof of Lemma \ref{lemma:tech}}

In this appendix, we prove the two formulae \eqref{acon-comm1} and \eqref{acon-comm2} for the completely contracted matrix elements $\langle A\rangle_{n,k}^{\rm con}$ \eqref{def-Acon}.

Recall that a contraction $C\in\CC_{n,k}$ is a set of pairs,
\begin{align}
 C=\{(l_1,r_1),...,(l_{|C|},r_{|C|})\}\,,
\end{align}
with $|C|\leq\min\{k,n-k\}$. The ``right indices'' satisfy $r_1,...,r_{|C|}\in\{1,...,k\}$, and the ``left indices'' $l_1,...,l_{|C|}\in\{k+1,...,n\}$. As before, we write $\bl_C$ and $\br_C$ for the sets $\{l_1,...,l_{|C|}\}$ and $\{r_1,...,r_{|C|}\}$, respectively.

We will need to distinguish between those contractions $C\in\CC_{n,k}$ which do not contract $k+1$, i.e. fulfill $k+1\notin\bl_C$, and those contractions which have $k+1\in\bl_C$ as a left index. The former set will be denoted $\hat{\CC}_{n,k}$, and the latter $\check{\CC}_{n,k}$. The set of all contractions is the disjoint union $\CC_{n,k}=\hat{\CC}_{n,k}\sqcup \check{\CC}_{n,k}$.

Also recall the shorthand notations $\delta_{l,r}:=\delta(\te_l-\te_r)$, $S_{a,b}:=S_2(\te_a-\te_b)$ and the definitions of $\delta_C$ and $S_C^{(k)}$,
\begin{align}\label{def-dS-app}
\delta_C		&:=	(-1)^{|C|}\prod_{j=1}^{|C|} \delta_{l_j,r_j}
\,,\qquad 
S_C^{(k)}		:=	\prod_{j=1}^{|C|} \prod_{m_j=r_j+1}^{l_j-1}
				S^{(k)}_{m_j,r_j}
				\cdot
				\prod_{\genfrac{}{}{0pt}{}{r_i<r_j}{l_i < l_j}}
				S^{(k)}_{r_j,l_i}
				\,,\\
S^{(k)}_{a,b}
			&:=
			\left\{
			\begin{array}{lll}
			S_{b,a}	&\;;\;	& a\leq k<b\;\;{\rm or}\;\; b\leq k< a\\
			S_{a,b}	&\;;\;	& {\rm otherwise}
			\end{array}
			\right.
\,.
\end{align}

Note that a contraction $C'\in\check{\CC}_{n,k}$ is always a union $C'=C\cup\{(k+1,r)\}$, where $C\in\hat{\CC}_{n,k}$ has length $|C|=|C'|-1$, and $r\notin \br_C$. In this situation, there holds
\begin{align}
\delta_{C'} &= -\delta_{k+1,r}\cdot\delta_C\,, 
\\
S_{C'}^{(k)}	&=
				\prod_{j=1}^{|C|} \prod_{m_j=r_j+1}^{l_j-1}
				S^{(k)}_{m_j,r_j}
				\cdot
				\prod_{m=r+1}^k S^{(k)}_{m,r}
				\cdot
				\prod_{\genfrac{}{}{0pt}{}{r_i<r_j}{l_i < l_j}}
				S^{(k)}_{r_j,l_i}
				\cdot
				\prod_{\genfrac{}{}{0pt}{}{r_i < r}{l_i<k+1}} S^{(k)}_{r,l_i}
				\cdot
				\prod_{\genfrac{}{}{0pt}{}{r<r_j}{k+1<l_j}}S^{(k)}_{r_j,k+1}
				\nonumber\\
				&=
				S_C^{(k)} \cdot \prod_{m=r+1}^k S_{m,r} \cdot \prod_{r<r_j}S_{k+1,r_j}\,, 
\end{align}
since $l_1,...,l_{|C|}>k+1$. Taking into account $S_{a,b}={S_{b,a}}^{-1}$ \eqref{s2rel}, we get
\begin{align}\label{congehampel1}
\delta_{C'}\cdot S_{C'}^{(k)}
&=
-\delta_C\cdot S_C^{(k)}\cdot \delta_{k+1,r}\cdot\prod_{\genfrac{}{}{0pt}{}{m=r+1}{m\neq r_j\,{\rm for}\, r_j>r}}^k S_{m,k+1}\,.
\end{align}

Similarly, contractions $C''\in\check{\CC}_{n,k+1}$ contracting $k+1$ (as a {\em right} index) are unions of the form $C''=\{(l,k+1)\}\cup C$, with $C\in\hat{\CC}_{n,k}$ and $l\notin\bl_C$. By a computation analogous to the one above one finds in this situation
\begin{align}\label{congehampel2}
\delta_{C''}&=-\delta_C\cdot\delta_{l,k+1}
\,,\qquad
S_{C''}^{(k+1)}
=
S_C^{(k+1)}\cdot\!\!\!\!\prod_{\genfrac{}{}{0pt}{}{m=k+2}{m\neq l_i\,{\rm for}\, l_i<l}}^{l-1} S_{k+1,m}
\,.
\end{align}

Now consider some contraction $C\in\hat{\CC}_{n,k}$. Repeated application of the relations of Zamolodchikov's algebra \eqref{ZFdis} yields (cf. \eqref{def-arl})
\begin{align}
\langle\bl_C|\,A\,|\br_C\rangle_{n,k}
=&\langle
\zd_{k+2}\cdots\widehat{\zd_{l_1}}\cdots\widehat{\zd_{l_{|C|}}}\cdots\zd_n\Om
\,,z_{k+1}A\,
\zd_k\cdots\widehat{\zd_{r_1}}\cdots\widehat{\zd_{r_{|C|}}}\cdots\zd_1\Om
\rangle\nonumber
\\
=&\;
\langle
\bl_C \cup \{k+1\}
|
\,[z_{k+1},A]\,
|
\br_C
\rangle_{n,k}\label{z-kalk-1}
\\
&+
\sum_{\genfrac{}{}{0pt}{}{r=1}{r\notin \sbr_C}}^k \delta_{k+1,r}
\!\!\!\!
\prod_{\genfrac{}{}{0pt}{}{m=r+1}{m\neq r_j\,{\rm for}\, r_j>r}}^k
\!\!\!\!
S_{m,k+1} 
\cdot
\langle
\bl_C \cup \{k+1\}
|\,A\,|
\br_C\cup\{r\}
\rangle_{n,k}
\,.\nonumber
\end{align}
Consider the last line, multiplied with $\delta_C S_C^{(k)}$ and summed over all $C\in\hat{\CC}_{n,k}$. Taking into account the remarks made at the beginning of the proof, there holds $\sum_{C'\in\check{\CC}_{n,k}}=\sum_{r=1,r\notin\sbr_C}^k\sum_{C\in\hat{\CC}_{n,k}}$, with the contractions $C$ and $C'$ being related by $C'=C\cup\{(k+1,r)\}$. Moreover, the delta distributions and scattering functions appearing in \eqref{z-kalk-1} are the same as in \eqref{congehampel1}. So we conclude
\begin{align*}
\sum_{C\in\hat{\CC}_{n,k}}
\delta_C S_C^{(k)}
\langle \bl_C|\,A\,|\br_C\rangle_{n,k}
=&
\sum_{C\in\hat{\CC}_{n,k}}
\delta_C S_C^{(k)}
\langle \bl_C\cup\{k+1\}|\,[z_{k+1},A]\,|\br_C\rangle_{n,k}
\\
&- \sum_{C'\in\check{\CC}_{n,k}}
\delta_{C'} S_{C'}^{(k)}
\langle \bl_{C'}|\,A\,|\br_{C'}\rangle_{n,k}
\,,
\end{align*}
and as $\CC_{n,k}=\hat{\CC}_{n,k}\sqcup \check{\CC}_{n,k}$,
\begin{align*}
\sum_{C\in\hat{\CC}_{n,k}}
\delta_C S_C^{(k)}
\langle \bl_C\cup\{k+1\}|\,[z_{k+1},A]\,|\br_C\rangle_{n,k}
=
\sum_{C\in\CC_{n,k}}
\delta_C S_C^{(k)}
\langle \bl_C|\,A\,|\br_C\rangle_{n,k}
\,.
\end{align*}
Since the right hand side coincides with $\langle A\rangle^{\rm con}_{n,k}$ \eqref{def-Acon}, this proves the first formula \eqref{acon-comm1} of Lemma \ref{lemma:tech}.
\\
\\
For the second formula \eqref{acon-comm2}, we argue in a similar manner. Considering a contraction $C\in\hat{\CC}_{n,k}$, the relations \eqref{ZFdis} of Zamolodchikov's algebra and $\overline{S_{a,b}}=S_{b,a}$ \eqref{s2rel} imply
\begin{align}
\langle \bl_C\cup&\{k+1\}|\,[A,\zd_{k+1}]\,|\br_C\rangle_{n,k}
=
\langle \bl_C|\,A\,|\br_C\rangle_{n,k+1}
\label{refforend}
\\
&\qquad-
\sum_{\genfrac{}{}{0pt}{}{l=k+2}{l\notin\sbl_C}}^n
\delta_{l,k+1}\!\!\!\!
\prod_{\genfrac{}{}{0pt}{}{m=k+2}{m\neq l_i\,{\rm for}\,l_i<l}}^{l-1}
\!\!\!\!
S_{k+1,m}
\cdot \langle \bl_C\cup\{l\}|\,A\,|\br_C\cup\{k+1\}\rangle_{n,k+1}
\,.\nonumber
\end{align}
According to the remarks made at the beginning of the proof, all contractions in $\check{\CC}_{n,k+1}$ are of the form $C'':=C\cup\{(l,k+1)\}$, $C\in\hat{\CC}_{n,k}$, $l\notin\bl_C$, i.e. we have the equality of sums  $\sum_{l=k+2,l\notin\sbl_C}^n\sum_{C\in\hat{\CC}_{n,k}}=\sum_{C''\in\check{\CC}_{n,k+1}}$. Taking into account the relations \eqref{congehampel2}, it follows that the second term on the right hand side in \eqref{refforend}, multiplied with $\delta_C S_C^{(k+1)}$ and summed over $C\in\hat{\CC}_{n,k}$, gives $\sum_{C''\in\check{\CC}_{n,k+1}}\delta_{C''}S_{C''}^{(k+1)}\langle\bl_{C''}|\,A\,|\br_{C''}\rangle_{n,k+1}$. As the first term in \eqref{refforend} yields the sum over all $C\in\hat{\CC}_{n,k}$, and since $\CC_{n,k+1}=\hat{\CC}_{n,k}\sqcup \check{\CC}_{n,k+1}$, we arrive at
\begin{align*}
\sum_{C\in\hat{\CC}_{n,k}}
\delta_C S_C^{(k+1)}
\langle \bl_C\cup\{k+1\}&|\,[A,\zd_{k+1}]\,|\br_C\rangle_{n,k}
\\
&=
\sum_{C\in\CC_{n,k+1}}
\delta_C S_C^{(k+1)}
\langle \bl_C|\,A\,|\br_C\rangle_{n,k+1}
=
\langle A\rangle_{n,k+1}^{\rm con}\,.
\end{align*}
This is the desired equation \eqref{acon-comm2}.{\hfill $\square$ \\[2mm] \indent}


\end{document}